\documentclass[leqno]{LMCS}
\usepackage[british]{babel}
\usepackage{amsmath}
\usepackage{amssymb}
\usepackage{xspace}
\usepackage{url}
\usepackage{multicol}
\usepackage{enumerate,hyperref}
\newcommand{\mathcmd}[1]{\ensuremath{#1}\xspace}
\makeatletter
\def\@cnt#1{cnt#1}
\def\@lbl#1{lbl#1}
\def\@fmt#1{lblfmt#1}
\def\@wdt#1{maxlbl#1}
\def\@mark#1{mark#1}
\def\@leftlbldelim#1{leftlbldelim#1}
\def\@rightlbldelim#1{rightlbldelim#1}
\def\@midlbldelim#1{midlbldelim#1}
\def\@prelbl#1{prelbl#1}
\def\@postlbl#1{postlbl#1}
\def\@preprelbl#1{preprelbl#1}
\def\@postpostlbl#1{postpostlbl#1}
\def\@parentlblfmt#1{parentlblfmt#1}
\def\@usercmd#1{usercmd#1}
\def\@@fmt#1#2{#1\arabic{\@cnt{#2}}}
\def\@@parentfmt#1{\@nameuse{\@lbl{#1}}}
\def\reusecounter#1{\@nmbrlisttrue\def\@listctr{#1}}
\def\@mrklist#1#2{%
\@namedef{\@mark{#2}}{#1}
\@namedef{\@leftlbldelim{#2}}{(}
\@namedef{\@rightlbldelim{#2}}{)}
\@namedef{\@prelbl{#2}}{}
\@namedef{\@postlbl{#2}}{}
\@namedef{\@preprelbl{#2}}{}
\@namedef{\@postpostlbl{#2}}{}
\@namedef{\@fmt{#2}}{\@@fmt}
\@namedef{\@usercmd{#2}}{\usecounter{\@cnt{#2}}}%
\@ifnextchar[{\@ymrklist{#2}{#1}}{\@xmrklist{#2}{#1}}%
}
\def\@xmrklist#1#2{%
\newcounter{\@cnt{#1}}
\@namedef{\@lbl{#1}}{\@nameuse{\@fmt{#1}}{#2}{#1}}
\df@mrklist{#1}{#2}
}
\def\df@mrklist#1#2{%
\@namedef{the\@cnt{#1}}{%
\@nameuse{\@leftlbldelim{#1}}\@nameuse{\@lbl{#1}}\@nameuse{\@rightlbldelim{#1}}}
\@namedef{\@wdt{#1}}{\@nameuse{the\@cnt{#1}}}
\newenvironment{#1}%
{%
 \begin{list}%
        {\@nameuse{\@preprelbl{#1}}\@nameuse{the\@cnt{#1}}\@nameuse{\@postpostlbl{#1}}}%
        {\setlength{\itemsep}{0pt}
         \setlength{\itemindent}{0pt}
         \setlength{\leftmargin}{0pt}
         \settowidth{\labelwidth}{%
                \@nameuse{\@preprelbl{#1}}%
                \@nameuse{\@prelbl{#1}}%
                \@nameuse{\@wdt{#1}}%
                \@nameuse{\@postlbl{#1}}%
                \@nameuse{\@postpostlbl{#1}}%
         }
         \leftmargin\labelwidth
         \advance\leftmargin\labelsep
         \@nameuse{\@usercmd{#1}}%
         \def\makelabel####1{\hss\llap{\@nameuse{\@prelbl{#1}}\ignorespaces ####1\@nameuse{\@postlbl{#1}}}}%
         \let\@olditem=\@item
                   \def\@item[####1]{%
                    \@olditem[####1]%
 \protected@edef\@currentlabel{####1}%
            }
        }
}%
{\let\@item=\@olditem\end{list}}
}
\def\@ymrklist#1#2[#3]{%
\@ifundefined{end#3}%
{%
\newcounter{\@cnt{#1}}[#3]
\@namedef{\@parentlblfmt{#1}}{\@nameuse{the#3}}
}%
{%
\newcounter{\@cnt{#1}}[\@cnt{#3}]
\@namedef{\@parentlblfmt{#1}}{\@@parentfmt{#3}}
}%
\@namedef{\@midlbldelim{#1}}{.}
\@namedef{\@lbl{#1}}{%
        \@nameuse{\@parentlblfmt{#1}}%
        \@nameuse{\@midlbldelim{#1}}\@nameuse{\@fmt{#1}}{#2}{#1}}
\df@mrklist{#1}{#2}
}
\newcommand\newmrklist[1][]{\@mrklist{#1}}
\makeatother
\newmrklist[wd]{wdlist}

\newmrklist[n]{nlist}
\newmrklist[t]{tlist}
\newmrklist[m]{mlist}
\newcommand{\mlfont}{\textsf}
\newcommand{\ml}[1]{\mathcmd{\mlfont{\upshape #1}}}
\newcommand{\IPC}{\ml{IPC}}
\newcommand{\KFour}{\ml{K4}}
\newcommand{\SFive}{\ml{S5}}
\newcommand{\dlfont}{\mathcal}
\newcommand{\dl}[1]{\mathcmd{\dlfont{#1}}}
\newcommand{\ALC}{\dl{ALC}}
\newcommand{\ALCO}{\dl{ALCO}}
\newcommand{\ALBO}{\dl{ALBO}}
\newcommand{\SHOIQ}{\dl{SHOIQ}}
\def\lAnd{\wedge}
\def\lOr{\vee}
\def\lImp{\rightarrow}
\def\lNot{\neg}

\let\And\dAnd
\let\Or\dOr
\let\Imp\dImp
\let\Not\dNot

\newcommand{\sub}{\mathcmd{\textsf{\upshape sub}}}

\newcommand{\metasmbfont}{\mathsf}
\newcommand{\lang}[1]{\mathcmd{\mathcal{#1}}}
\newcommand{\Lang}{\mathcal}
\newcommand{\seq}[1]{\overline{#1}}
\newcommand{\ecl}[1]{\|#1\|}
\makeatletter
\def\@define#1{%
    \mathcmd{%
        \stackrel%
            {\mbox{%
                \tiny\ensuremath{\metasmbfont{def}}%
                  }%
            }%
            {{#1}}%
            }%
              }
\def\sm@shdefine#1{%
\smash[t]{\@define{#1}}%
}
\newcommand{\define}{%
        \mathchoice{\@define{\ =\ }}{\sm@shdefine{=}}{\sm@shdefine{=}}{\sm@shdefine{=}}%
}
\newcommand{\defiff}{%
        \mathchoice%
            {\@define{\ \Longleftrightarrow\ }}%
            {\sm@shdefine{\Longleftrightarrow}}%
            {\sm@shdefine{\Leftrightarrow}}%
            {\sm@shdefine{\Leftrightarrow}}%
}
\makeatother
\newcommand{\tableaulblfont}{\sffamily\scriptsize}
\newcommand{\tableauleftlbldelim}{(}
\newcommand{\tableaurightlbldelim}{)}
\newcommand{\tand}{,\ \ }
\newcommand{\tor}{\,\mid\,}

\newcommand{\branch}[1]{\seg{#1}}
\newcommand{\seg}[1]{\mathcmd{\mathcal{#1}}}
\makeatletter
\def\tableaulblfmt#1{\text{\tableaulblfont\tableauleftlbldelim #1\tableaurightlbldelim}}
\def\@xtableaurule#1#2{%
\genfrac{}{}{}{0}{#1}{#2}%
}
\def\@ytableaurule#1#2[#3]{%
\def\lbl@{#3}
\iftagsleft@%
\tableaulblfmt{#3}\@xtableaurule{#1}{#2}%
\else%
\@xtableaurule{#1}{#2}\tableaulblfmt{#3}%
\fi%
\let\label\ltx@label
\let\@currentlabel\lbl@
}
\def\tableaurule#1#2{%
\@ifnextchar[{\@ytableaurule{#1}{#2}}{\@xtableaurule{#1}{#2}}%
}
\makeatother
\newcommand{\proverfont}{\textsc}
\newcommand{\mettel}{\mathcmd{\proverfont{MetTeL}}}
\newcommand{\lotrec}{\mathcmd{\proverfont{LoTREC}}}
\newcommand{\SO}{\dl{SO}}
\newcommand{\inverse}{\mathord{\sim}}
\newcommand{\FO}{\textsf{\slshape FO}\xspace\/}
\newcommand{\Conn}{\textsf{\slshape Conn}\xspace\/}
\newcommand{\Sorts}{\textsf{\slshape Sorts}\xspace\/}
\newcommand{\range}{{\upharpoonright}}
\newcommand{\I}{\mathcal{I}}
\newcommand{\simB}{\mathop{\sim_\branch{B}}}
\newcommand{\IB}{{\I(\branch{B})}}
\newcommand{\Rf}{\mathcmd{\mathop{\textsf{\upshape ref}}}}
\newcommand{\Tr}{\mathcmd{\mathop{\textsf{\upshape tr}}}}
\newcommand{\Rfr}{\Rf}
\newcommand{\Rfe}{\Tr}
\theoremstyle{plain}

\newtheorem{theorem}[thm]{Theorem}
\newtheorem{lemma}[thm]{Lemma}
\newtheorem{corollary}[thm]{Corollary}

\def\doi{7 (2:6) 2011}
\lmcsheading%
{\doi}
{1--32}
{}
{}
{May\phantom{.}~29, 2010}
{May\phantom{.~0}5, 2011}
{}

\begin{document}
\title{Automated Synthesis of Tableau Calculi\rsuper*}%

\author[R.~A.~Schmidt]{Renate A.\ Schmidt}
\address{School of Computer Science, The University of Manchester, UK}
\email{\{renate.schmidt,dmitry.tishkovsky\}@manchester.ac.uk}

\author[D.~Tishkovsky]{Dmitry Tishkovsky}
\address{\vskip-6 pt}


\titlecomment{{\lsuper*}The paper is an extended and improved version of~\cite{SchmidtTishkovsky-ASTC-2009}.}

\begin{abstract}
This paper presents a method for synthesising sound and complete
tableau calculi. 
Given a specification of the formal semantics of a logic, the method
generates a set of tableau inference rules that can then be
used to reason within the logic.
The method guarantees that the generated rules 
form a calculus which is sound and constructively complete.
If the logic can be shown to admit finite filtration with respect to
a well-defined first-order semantics
then adding a general blocking mechanism provides a terminating
tableau calculus. 
The process of generating tableau rules can be completely automated
and produces, together with the blocking mechanism,
an automated procedure for generating tableau decision procedures.
For illustration we show the workability of the approach for
a description logic with transitive roles and
propositional intuitionistic logic. 
\end{abstract}

\maketitle

\section{Introduction}

Tableau-based reasoning is popular in many areas of computer
science and various branches of logic.
For description logics and ontology reasoning  
they provide the
main method for doing reasoning (see, for example,~\cite{BaaderSattler01,HorrocksSattler-TDP_SHOIQ-2007}; some
recent work is~\cite{SchmidtTishkovsky-UTD+-2007,MotikShearerHorrocks09}).
For modal logics and applications such as multi-agent
systems tableau approaches are frequently
used (see, for example,~\cite{Fitting-PMMIL-1983,%
CastilhoFarinasDelCerroGasquetHerzig97,Gore-TMMTL-1999,Massacci00a,%
CialdeaMayerCerrito01,HorrocksHustadtSattlerSchmidt07}; some recent
work is~\cite{GorankoShkatov09,BalbianiEtAl10}).
Tableau calculi have been developed and are being
used for non-classical logics such as intuitionistic
logic~\cite{Fitting-PMMIL-1983,AvelloneEtAl97}, conditional
logic~\cite{DBLP:conf/csl/AlendaOST10}, logics of metric and
topology~\cite{HustadtTishkovskyWolterZakharyaschev-ARMT-2006}
and hybrid
logics~\cite{Tzakova99,BolanderBlackburn07,CialdeaMayerCerrito10}.
Rather than developing tableau calculi one by one for individual logics,
it is possible to develop tableau calculi in a systematic way for
large classes of logics.
This is evident from the literature in all these areas and studies such
as~\cite{Heuerding98,FarinasDelCerroGasquet02,AvelloneEtAl97}.

In this paper we want to go further and investigate the
possibility to generate tableau calculi automatically from the
specification of a logic.
We assume that the logic of interest is defined by a
high-level specification of its formal semantics. 
Our aim is to turn this into a set of inference rules that provides
a sound and complete deduction calculus for the logic.
Ideally we also want to be able to guarantee termination if the logic
is decidable.
Automated synthesis of calculi is a challenging problem and
in general it is of course not possible to turn every specification
of a logic into a sound, complete and terminating deduction calculus.
It is however possible to describe classes of logical specifications
for which the problem is solvable uniformly.

In previous work we have shown that it is possible to synthesise
tableau calculi for modal logics by translation to first-order logic
combined with first-order resolution~\cite{Schmidt06b}.
In this approach the semantic specification of a logic is transformed
into clausal form and then a set of inference rules.
Soundness and completeness of the generated calculus follows from the
soundness and completeness of the simulating resolution refinement
used.
In the present paper we introduce another approach for generating tableau calculi.
Rather than proceeding via simulation by resolution, 
our approach generates a tableau calculus 
directly from the specification of a logic.
For traditional modal logics essentially the same tableau calculi can be obtained,
but for more expressive dynamic modal logics
and description logics the method in~\cite{Schmidt06b} produces calculi
with introduction rules, whereas the method in this paper can
be used to produce calculi with only elimination
rules.

In other previous work we have described a framework for
turning sound and complete tableau calculi into decision
procedures~\cite{SchmidtTishkovsky-GTM+-2008}.
The key for this framework is the unrestricted blocking mechanism
from~\cite{SchmidtTishkovsky-UTD+-2007} which is added to 
the given calculus in order to turn it into a terminating calculus.
Enhancing  
a tableau calculus 
with the unrestricted blocking mechanism produces a terminating tableau
calculus, whenever the logic can be shown to admit finite filtration
with respect to its semantics~\cite{SchmidtTishkovsky-GTM+-2008}.
More specifically, the prerequisites  
are that the following
conditions all hold.
\begin{enumerate}[(1)]
\item
\label{efmp_item}
The logic admits the effective finite model property shown by a
filtration argument.
\item
\label{sound_constructively_complete_item}
The tableau calculus is sound and constructively complete.
\item
\label{subexpression_property_item}
A weak form of subexpression property holds for tableau
derivations.
\end{enumerate}
Constructive completeness is a slightly stronger notion than
completeness and means that for every open branch in a tableau there
is a model which reflects all the expressions (formulae)
occurring on the branch.
The subexpression property says that every expression 
in a derivation  
is a subexpression of the input expression
with respect to a finite subexpression closure operator.

In order to be able to exploit this `termination through blocking'
result from~\cite{SchmidtTishkovsky-GTM+-2008}, in this paper, our goal is to
synthesise tableau calculi that satisfy the
prerequisites~(\ref{sound_constructively_complete_item})
and~(\ref{subexpression_property_item}).
It turns out that, provided
the specification of the semantics of the logic is
well-defined in a certain sense, the subexpression property can be
imposed on the generated calculus.
Crucial is the separation of the syntax of the logic from the
`extras' in the meta-language needed for the semantic specification of
the logic.
The process of
generating tableau calculi can be completely automated and gives, together
with the unrestricted blocking mechanism and the results
in~\cite{SchmidtTishkovsky-UTD+-2007,SchmidtTishkovsky-GTM+-2008}, an
automated procedure for generating tableau decision procedures for
logics, whenever they have the effective finite model property with
respect to a well-defined first-order semantics, that is, condition~\eqref{efmp_item} holds.

The tableau synthesis method introduced in this paper works as follows.
The user defines the formal semantics of the given logic in a
many-sorted first-order language so that certain well-definedness
conditions hold.
The semantic specification of the logic is then automatically reduced
to Skolemised implicational forms which are further transformed into tableau
inference rules.
Combined with a set of default closure and equality rules, the
generated rules provide a sound and constructively complete calculus
for the logic.
Under certain conditions the set of rules can be further
refined.
If the logic can be shown to admit finite filtration,
then the generated calculus can be automatically turned into
a terminating calculus by adding the unrestricted blocking 
mechanism from~\cite{SchmidtTishkovsky-UTD+-2007}.

The method is intended to be as general as possible, and cover
as many logics as possible.
Our main focus is non-classical logics and description
logics. 
As case studies we consider the
application of the method to propositional intuitionistic logic
\IPC~\cite{Kripke-SAIL1-1965} and the
description logic \SO. 
Propositional intuitionistic logic provides a nearly perfect example 
because the semantics of the logical connectives
is  not Boolean and the semantics is restricted by
a background theory.
In addition, the logic is simple.
\SO is the extension of the description logic \ALC with 
singleton
concepts (or nominals) and transitive roles.
\SO~is a fragment of many expressive description logics
considered in the
literature~\cite{BaaderCalvaneseEtal03} and
is the analogue of the hybrid~\cite{Blackburn-WHL-1998} version of the standard modal logic \KFour~\cite{BlackburnDeRijkeVenema01}.

The paper is structured as follows.
Section~\ref{section_semantic_specification} defines the apparatus
for specifying the logic of interest.
It consists of two languages, a language for specifying
the syntax of the logic and a language for specifying its semantics.
How to specify the semantics of a logic 
is described in Section~\ref{section_semantics}.
Because there are many ways of writing semantic specifications, 
in this paper, we focus on what we call well-defined
semantic specifications for which sound and complete tableau
calculi can be generated.
The tableau generation process is presented
in Section~\ref{section_tableau_synthesis}, and
Section~\ref{section_sound_constructively_complete} proves soundness and
constructive completeness of the generated calculus.
Sections~\ref{section_optimisation} and~\ref{section: refinement 2}
discuss two techniques for refining a calculus.
The first refinement aims at 
reducing branching in derivations. The second refinement aims at reducing the use of
extraneous constructs in the language of the tableau calculus.
In Section~\ref{section: blocking} we show how the unrestricted
blocking mechanism of \cite{SchmidtTishkovsky-UTD+-2007} can be used to
obtain terminating tableau calculi for logics with the effective finite
model property. 
To illustrate the approach we use the description logic \SO as
a running example throughout the paper.
In Section~\ref{section_case_studies} the approach is applied to
propositional intuitionistic logic.
The paper concludes with a discussion of the approach.

The paper is written using terminology of description logics, but
all the results apply equally to modal logics and other non-classical
logics.
In most cases where we use the word `expression' we could have
equally used the words `formula' or `logical term'. 

\section{The Specification Languages}
  \label{section_semantic_specification}
  
In order for the user to specify the semantics of the given logic for
which they want to develop a tableau calculus there are two specification
languages: 
\begin{enumerate}[(1)]
\item
an object language for defining the syntax
of the logic, and 
\item
a meta-language for specifying the semantics of
the logic.
\end{enumerate}

For the sake of generality the \emph{object language}, denoted by
$\Lang{L}$, is a many-sorted propositional language, thus allowing
for the specification of many-sorted propositional
logics including modal logics, description logics and other
non-classical logics.

Throughout the paper the standard notation $\omega$ 
is used for the smallest infinite countable ordinal,
that is, $\omega=\{0,1,2,\ldots\}$.

Let $\Sorts\define\{0,1,\ldots,N\}$ be the index set of
the sorts of the object language.
The idea is that, for $n=1,\ldots,N$, symbols of sort~$n$ are interpreted as $n$-ary
relations
and symbols of sort~$0$ are interpreted as domain elements.
Of the sorts, the sort~$1$ is regarded as the \emph{primary sort}.

Let $\Conn$ 
be a countable set of the logical connectives of the logic to be
specified.
Every connective~$\sigma$ in $\Conn$ is associated
with a tuple 
$(i_1,i_2,\ldots,i_{m+1}) \in \Sorts^{(m+1)}$, 
where~$m\geq 0$.
The last argument~$i_{m+1}$ is the sort of the expression obtained
by applying $\sigma$ to expressions of sorts $i_1,i_2,\ldots,i_{m}$,
respectively.
We say that $\sigma$ is an $m$-ary connective of sort
$(i_1,i_2,\ldots,i_{m+1})$.

The object language $\Lang{L}$ is defined over
an alphabet given by a set of sorts $\Sorts$, a set of connectives
$\Conn$, a countable set of variable symbols $\{p^i_j\mid
i\in\Sorts, j\in\omega\}$, and a countable set of 
constant symbols $\{q^i_j\mid i\in\Sorts, j\in\omega\}$.
$\Lang{L}$ is defined as the set of \emph{expressions} over the 
alphabet closed under the connectives in $\Conn$.
More formally,
let $\Lang{L}\define
\bigcup_{i\in\Sorts}\Lang{L}^i$, where
each $\Lang{L}^i$ denotes the \emph{set of expressions of sort~$i$} 
defined as the smallest set of expressions satisfying
the following conditions:
\begin{enumerate}[$\bullet$]
 \item All variables $p^i_j$ and all constants
       $q^i_j$ in the alphabet are expressions belonging to $\Lang{L}^i$.
 \item For every connective $\sigma\in\Conn$ of sort
       $(i_1,i_2\ldots,i_{m+1})$, 
       $\sigma(E_1,\ldots,E_m)$ is an expression belonging to $\Lang{L}^{i_{m+1}}$,
       whenever $E_1,\ldots,E_m$ belong to
       $\Lang{L}^{i_{1}}, \ldots, \Lang{L}^{i_{m}}$, respectively.
\end{enumerate}\medskip

\noindent Symbols, expressions and connectives in the language $\Lang{L}$ are also referred to
as $\Lang{L}$-symbols, $\Lang{L}$-expressions and $\Lang{L}$-connectives.
Variables and constants in $\Lang{L}$ are called \emph{atomic} $\Lang{L}$-expressions.
We refer to expressions in $\Lang{L}^0$ as \emph{individuals},
expressions in $\Lang{L}^1$ as \emph{concepts},
and expressions in $\Lang{L}^2$ as \emph{roles}.
That is, individuals are expressions of sort~0, concepts are expressions
(or formulae) of the primary sort and roles are expressions
(or formulae) of sort~2.

For an $\Lang{L}$-expression $E$, 
the notation $E(p_1,\ldots,p_m)$ indicates that $p_1,\ldots,p_m$
are (distinct) variables occurring in the expression $E$.
To avoid ambiguity in this notation
we standardly assume that all the variables of the language $\Lang{L}$
are linearly ordered by an ordering $<_v$ and $p_1<_v\cdots<_v p_m$.
$E(E_1,\ldots,E_m)$ denotes the expression obtained by uniformly
substituting $E_i$ into $p_i$, for all $i=1,\ldots,m$.
Similarly, if $X$ is a set of
$\Lang{L}$-expressions depending on variables $p_1,\ldots,p_m$,
we indicate this as $X(p_1,\ldots,p_m)$ and
denote by $X(E_1,\ldots,E_m)$ the set of expressions
which are instances of expressions from $X$
under uniform substitution of the expressions $E_1,\ldots,E_m$
into $p_1,\ldots,p_m$, respectively.

Throughout the paper we use the logic~\SO as a running example. 
Recall that \SO is the description logic \ALC extended with nominals, or singleton
concepts, and transitive roles.

The object language $\Lang{L}_\SO$ for specifying the syntax of
$\SO$ consists of three sorts, namely~$0$, $1$ and~$2$ for individuals,
concepts, and roles, respectively.
Atomic expressions of sort $0$ are individual variables from a
countable set $\{p^0_j\mid j\in\omega\}$.
We denote individual variables also by $\ell_0,\ell_1,\ldots$.
The variables $p^1_j$ are of sort~$1$ and are the concept symbols.
In this paper concept symbols are denoted by $p_0,p_1,\ldots$.
The variables  $p^2_j$ of sort $2$ are the atomic roles and are
denoted by $r_0,r_1,\ldots$.

The connectives in $\Lang{L}_\SO$ are the following:
\begin{enumerate}[$\bullet$]
 \item The `singleton concept' connective $\{\cdot\}$ of the sort $(0,1)$.
    That is, $\{\ell\}$ is a concept for every individual $\ell$.
 \item The Boolean connectives $\Or$ and $\Not$ of sorts $(1,1,1)$ and $(1,1)$, respectively.
       As usual, we use infix notation for $\Or$ and prefix notation for $\Not$.
       Thus, $C\Or D$ and $\Not C$ are concepts for any concept expressions $C$ and $D$.
 \item The existential restriction connective $\exists\cdot.\cdot$ of sort $(2,1,1)$.
    That is, $\exists r.C$ is a concept for any role expression $r$ and concept expression $C$.
\end{enumerate}\medskip

\noindent Thus, expressions of $\Lang{L}_\SO$ are built from
individuals, concept symbols and role symbols using the given
connectives, and there are no other expressions in $\Lang{L}_\SO$.
In this language, individual and role expressions are allowed
to be atomic only.

\bigskip
The \emph{meta-language} in which the semantics of the given logic
is specified is a many-sorted first-order language with
equality and is denoted by $\FO(\Lang{L})$.
$\FO(\Lang{L})$ extends the object language $\Lang{L}$,
the idea being that $\Lang{L}$-expressions are represented as terms in $\FO(\Lang{L})$
and $\Lang{L}$-connectives as functions.

Formally, $\FO(\Lang{L})$ is defined as an extension of $\Lang{L}$ with one additional
sort, namely \text{$N+1$}, additional symbols, the standard first-order connectives
$\lNot$, $\lOr$, $\lAnd$, $\lImp$, the equivalence connective $\equiv$,
first-order quantifiers $\exists$ and $\forall$, and the equality predicate $\approx$.
Thus the sorts of $\FO(\Lang{L})$ are $\Sorts \cup \{ N+1 \} = \{ 0, \ldots,
N, N+1\}$. We call the additional sort $N+1$ the \emph{domain sort}, 
and symbols over this sort are called the \emph{domain
symbols}.
The additional symbols comprise of a countable set of variable
symbols $\{x,y,z,x_0,y_0,z_0,\ldots\}$ of the domain sort,
a countable set of constants $\{a,b,c,a_0,b_0,c_0,\ldots\}$ of the
domain sort, 
function symbols $\{f,g,h,f_0,g_0,h_0,\ldots\}$ 
mapping argument terms
to terms of sort $N+1$, and
a countable set of constant predicate symbols
$\{P,Q,R,P_0,Q_0,R_0,\ldots\}$ of the domain sort (that is, argument
terms are required to be terms of sort $N+1$).
Intuitively, the domain sort contains symbols necessary for 
formalising semantic properties of the domain elements of 
interpretations of the target logic.

In addition, $\FO(\Lang{L})$ contains the symbols
$\nu_0,\ldots,\nu_N$, one for each sort in $\Sorts$ of
the object language.
In particular, $\nu_0$ is a unary function symbol
of sort $(0,N+1)$ (that is, a function from sort $0$ to sort $N+1$).
Each of the remaining~$\nu_i$ is a predicate symbol of
sort $(i,N+1,\ldots,N+1)$ with arity $i+1$.

The purpose of these symbols is to define the
semantics of the connectives of the logic by using conditions similar to satisfaction
conditions in standard definitions.
$\nu_0$ can be viewed as the interpretation mapping for individuals 
(represented as terms) in the object language.
All other $\nu_n$ can be viewed as interpretation mappings
for expressions in the object language;
they can be viewed as the `holds' or `satisfaction' predicates.

Finally, for every sort we assume the presence in $\FO(\Lang{L})$ of a binary
predicate symbol representing the equality predicate for that sort.
For reasons of simplicity, we use one symbol, namely $\approx$, for each
of the equality predicates.

\emph{Formulae} in $\FO(\Lang{L})$ are just first-order formulae defined over the symbols of $\FO(\Lang{L})$, where 
each expression in $\Lang{L}$ is represented
by a term in $\FO(\Lang{L})$.
In particular, each variable symbol $p^i_j$ in~$\Lang{L}^i$
is represented by a variable of sort $i$ in $\FO(\Lang{L})$,
each constant symbol $q^i_j$ in~$\Lang{L}^i$
is represented by a constant of sort $i$ in $\FO(\Lang{L})$,
and every connective~$\sigma$ is represented by 
a function of the same sort as $\sigma$.

To illustrate how expressions of a logic are represented in a
meta-language we continue our running example.
According to our definitions the meta-language $\FO(\Lang{L}_\SO)$
for \SO is a first-order language with sorts $0$, $1$, $2$ and $3$.
The interpretation symbols are $\nu_0$ (which is a function symbol) and
the holds predicate symbols $\nu_1$ and $\nu_2$.
Also included in $\FO(\Lang{L}_\SO)$ is the equality predicate symbol
$\approx$.

Every variable of $\Lang{L}_\SO$ is represented by a variable
of the corresponding sort in $\FO(\Lang{L}_\SO)$.
Thus, every individual variable $\ell$ in $\Lang{L}_\SO$ is represented
by a
variable of sort $0$ in $\FO(\Lang{L}_\SO)$.
Every concept symbol $p$ in $\Lang{L}_\SO$ is represented by a
variable of sort~$1$, and every role symbol~$r$ in
$\Lang{L}_\SO$ by a variable of sort~$2$
in $\FO(\Lang{L}_\SO)$.
Connectives of the object language become function symbols of an
appropriate sort in $\FO(\Lang{L}_\SO)$.
Thus, every expression in $\Lang{L}_\SO$ becomes a first-order term
of the corresponding sort.
For instance, the concept expression $\exists r.p$ is represented as
a term of sort $1$.

Whereas the sorts $0$, $1$, and $2$ are the sorts in the object
language $\Lang{L}_\SO$, the sort $3$ is a separate sort in
$\FO(\Lang{L}_\SO)$ with its own sets of variables, individual
constants, function symbols, and symbols of predicate constants.
Sort~3 is the domain sort for $\SO$.

Finally, for every individual $\ell$, $\nu_0(\ell)$ is a term of
sort~$3$ in $\FO(\Lang{L}_\SO)$, and $\nu_1(C,t)$ and $\nu_2(r,t,t')$
are atomic formulae  
of $\FO(\Lang{L}_\SO)$,
for any concept expression $C$, any role expression~$r$, and any terms
$t$ and $t'$ of sort $3$.

\bigskip
Before we describe how a logic can be defined in the meta-language
$\FO(\Lang{L})$ in the
next section, we fix some more notation and
terminology.
Let
$\seq{w}$ denote a sequence of first-order variables, that is
$\seq{w} \define w_1, \ldots, w_n$.
Similarly, let $\forall\overline{w}$ denote
the universal quantifier prefix on all variables $w_1,\ldots,w_n$,
that is, $\forall\seq{w}\define\forall w_1\cdots\forall w_n$.
For any set $S$ of formulae, $\forall S$~denotes
the universal closure of $S$, that is, the set 
\[
\forall S\define\{\forall\seq{w}\;\phi(\seq{w})\mid\phi(\seq{w})\in S\}.
\]
For every first-order formula $\psi$ we let 
\[
\inverse\psi\define\begin{cases}
                  \psi', & \text{provided $\psi=\Not\psi',$}\\
                  \Not\psi, &\text{otherwise.} 
                 \end{cases}
    \]

Formulae of $\FO(\Lang{L})$ in which all occurrences of the
$\Lang{L}$-variables $p^i_j$ (of sorts $i=0,\ldots,N$) are free
are called \emph{$\Lang{L}$-open} formulae.
An \emph{$\Lang{L}$-open sentence} is an $\Lang{L}$-open formula that
does \emph{not} have free occurrences of variables of the domain
sort $N+1$.

For example, the formula
\[
\forall y\,(\nu_1(\exists r.p,y)\lAnd\nu_2(r,x,y))
\]
is an $\Lang{L}_\SO$-open formula because the variables $p$ and $r$ occur
only freely.
Because the variable~$x$ of domain sort $3$ also occurs freely,
it is \emph{not} an $\Lang{L}_\SO$-open sentence.
In contrast, the formula
\[
\forall y\,(\nu_1(\exists r.p,y)\lAnd\forall x\,\nu_2(r,x,y))
\]
is an $\Lang{L}_\SO$-open sentence, because all the occurrences of the domain variables $x$
and $y$ are bound by quantifiers and all the occurrences of $p$ and $r$ are unbound.
The formulae
\[
\forall p\forall y\,(\nu_1(\exists r.p,y)\lAnd\nu_2(r,x,y))
\quad \text{and} \quad
\forall r\,(\nu_1(\exists r.p,y)\lAnd\nu_2(r,x,y))
\]
are not $\Lang{L}_\SO$-open because of the presence of quantified
variables of sorts other than the domain
sort ($p$ and $r$).
(The symbol $\exists$ in $\exists r.p$ should not be confused with
the existential quantifier of first-order logic.)

For any set~$S$ of $\Lang{L}$-open formulae in $\FO(\Lang{L})$ and
a set $X$ of $\Lang{L}$-expressions, let 
\begin{multline*}
 S\range X\define\{\phi(E_1,\ldots,E_m)\mid\phi(p_1,\ldots,p_m)\in S\ \text{and}\\
\text{all $\Lang{L}$-expressions occurring in $\phi(E_1,\ldots,E_m)$
belong to $X$}\}.
\end{multline*}
$S\range X$ is the set of instances of formulae
in $S$ under substitutions into the variables of~$\Lang{L}$ that do
not contain expressions outside $X$.

Suppose, for example,
\[
S\define\{\nu_1(\exists r.p,y),\nu_1(\Not p,x)\}
\quad \text{and} \quad
X\define\{r_0,p_0,p,p\And p_0,\exists r_0.p_0\}.
\]
Then the instantiations of formulae in $S$ relative to $X$
are
\begin{align*}
& \nu_1(\exists r_0.p_0,x), \ \nu_1(\exists r_0.p,x), \ 
\nu_1(\exists r_0.p\And p_0,x), \ \nu_1(\exists r_0.\exists r_0.p_0,x),
\\ &
\nu_1(\Not p_0,x), \ \nu_1(\Not p,x), \ \nu_1(\Not(p\And p_0),x), \ 
\nu_1(\Not\exists r_0.p_0,x).
\end{align*}
The only formula in this list where all $\lang{L}$-subexpressions
belong to $X$ is $\nu_1(\exists r_0.p_0,y)$.
Thus
\[
S\range X=\{\nu_1(\exists r_0.p_0,y)\}.
\]
The formula $\nu_1(\exists r_0.p,y)$ does not belong to $S\range X$ 
because $\exists r_0.p$ does not belong to $X$.
Other instances do not belong to $S\range X$ for similar reasons.

\section{Specifying the Semantics of an Object Language}
\label{section_semantics}

First, we define the model structures in terms of which the semantics of the object language is then defined.

An \emph{$\Lang{L}$-structure} is a tuple
$\I\define(\Lang{L}^0,\ldots,\Lang{L}^N,\Delta^\I,\nu_0^\I,\ldots,\nu_N^\I,a^\I,\ldots,P^\I,\ldots)$
where $\Delta^\I$ is a non-empty set,
$\nu_0(\ell)^\I\in\Delta^\I$ for every individual $\ell\in\Lang{L}^0$, 
$\nu_n^\I\subseteq\Lang{L}^n\times(\Delta^\I)^n$, for $0 < n \leq N$.
$a^\I \in \Delta^\I$ and
$P^\I\subseteq(\Delta^\I)^m$, where $m$ is the arity of~$P$.
For simplicity we omit the
sets $\Lang{L}^0,\ldots,\Lang{L}^N$ and simply write
\[\I=(\Delta^\I,\nu_0^\I,\ldots,\nu_N^\I,a^\I,\ldots,P^\I,\ldots).\]
Observe that an $\Lang{L}$-structure $\I$ is a first-order interpretation of the language $\FO(\Lang{L})$.

\medskip
For our sample logic \SO an $\Lang{L}_\SO$-structure is given by
a tuple $\I=(\Delta^\I,\nu_0^\I,\nu_1^\I,\nu_2^\I)$.
This means, the $\nu_i^\I$ are arbitrary interpretation functions for \SO-expressions.
As yet no additional conditions are assumed.
In the description logic literature
instead of a family of holds relations
$\nu_i$ just one holds relation $\nu$ is used,
resulting in the simpler and more familiar notation for an
interpretation, namely
$\I=(\Delta^\I,\nu^\I)$.

A \emph{valuation} in $\I$ is a mapping $\iota$ from the set of
variables and constants of~$\FO(\Lang{L})$ to $\Lang{L}\cup\Delta^\I$
such that $\iota(p^i_j),\iota(q^i_j)\in\Lang{L}^i$, and
$\iota(x_j),\iota(a_j)\in\Delta^\I$.
We use the standard notation $\I,\iota\models\phi$ to indicate 
a (first-order) formula $\phi$ is \emph{true} in the (first-order)
interpretation~$\I$ under valuation $\iota$.
Given a set of formulae $S$, we write $\I,\iota\models S$
if $\I,\iota\models\phi$ for every formula~$\phi$ in $S$.

We say that a valuation $\iota$
in an $\Lang{L}$-structure is \emph{canonical}
if every variable and constant of any sort $i=0,\ldots,N$
is mapped to itself, that is, $\iota(p^i_j)=p^i_j$ and
$\iota(q^i_j)=q^i_j$ for every variable $p^i_j$ and constant $q^i_j$
in the language $\Lang{L}$.
This means that the canonical valuation of any term of sort
$i=0,\ldots,N$
is the term itself.

It is not difficult to see that any
$\Lang{L}$-open formula $\phi$
is satisfiable in an $\Lang{L}$-structure
iff it is satisfiable in an $\Lang{L}$-structure
under a canonical valuation.

We write $S\models_c S'$ for sets of formulae $S$ and $S'$,
if, for every $\Lang{L}$-structure $\I$
and a canonical valuation $\iota$ in $\I$,
$\I,\iota\models S$ implies $\I,\iota\models S'$.
Similarly, we write $\I\models_c S$
iff there is a canonical valuation $\iota$ such that $\I,\iota\models S$.

Satisfiability for expressions of the
given logic is defined only for expressions of the primary sort,
that is, concept expressions.
We say 
a concept expression $C$ is \emph{satisfiable}
in~$\I$
if there is an element $a$ in $\Delta^\I$
such
that $(C,a)\in\nu_1^\I$, or equivalently, $\I\models_c\exists
x\;\nu_1(C,x)$. 
A concept expression $C$ is \emph{valid} in $\I$ if $\I\models_c\forall
x\;\nu_1(C,x)$.

\bigskip
Next we describe how the semantics of a given logic can be specified
in $\FO(\Lang{L})$, where~$\Lang{L}$ is the object language
of the logic. 

Let $S$ be any set of $\Lang{L}$-open sentences in $\FO(\Lang{L})$
and $\sigma$ be a connective of a sort $(i_1,\ldots,i_m,n)$.
A formula $\phi^\sigma$ in the language of $S$ \emph{defines
the connective $\sigma$} with respect to $S$ if
it does not contain~$\sigma$ and the following holds:
\begin{align}\label{eq: connective def}
\forall S\models\forall p_1\ldots\forall p_m\;\forall \seq{x}\;
(%
\nu_n(\sigma(p_1,\ldots,p_m),\seq{x})\equiv%
\phi^\sigma(p_1,\ldots,p_m,\seq{x})).%
\end{align}
Here 
$p_1,\ldots,p_m$ are variables of sorts $i_1,\ldots,i_m$ respectively.
If there is a formula $\phi^\sigma$ which defines
$\sigma$ with respect to $S$,
we also say $S$ \emph{defines}~$\sigma$ and
\[
\forall \seq{x}\;%
(%
\nu_n(\sigma(p_1,\ldots,p_m),\seq{x})\equiv%
\phi^\sigma(p_1,\ldots,p_m,\seq{x})),
\]
which is an $\Lang{L}$-open sentence, 
is a \emph{$\sigma$-definition with respect to $S$}.
Connective definitions are always $\Lang{L}$-open sentences,
that is, they do not contain any quantifiers over variables of sorts
$0,\ldots,N$ (these are implicitly regarded as being universally
quantified).

\begin{figure}[tb]
 \begin{gather*}
  \forall x\;(x\approx x)\qquad\quad \forall x\forall y\;(x\approx y\rightarrow y\approx x) \qquad\quad 
    \forall x\forall y\forall z\;(x\approx y\land y\approx z\rightarrow x\approx z)\\ 
  \forall x_1\cdots\forall x_n\forall y_i\;\left(P(x_1,\ldots,x_n)\land x_i\approx y_i\rightarrow P(x_1,\ldots x_{i-1},y_i,x_{i+1},x_n)\right)\\
  \forall p\,\forall x_1\cdots\forall x_n\forall y_i\;\left(\nu_n(p,x_1,\ldots,x_n)\land x_i\approx y_i\rightarrow\nu_n(p,x_1,\ldots x_{i-1},y_i,x_{i+1},x_n)\right)\\
 \begin{split}
  \forall p_1\cdots\forall p_m & \forall x_1\cdots\forall x_n\forall y_i\;
    (x_i\approx y_i\rightarrow\\
    &f(p_1,\ldots,p_m,x_1,\ldots,x_n)\approx f(p_1,\ldots,p_m,x_1,\ldots x_{i-1},y_i,x_{i+1},\ldots,x_n)%
    )
   \end{split}
 \end{gather*}
\caption{Default equality axioms in $\FO(\Lang{L})$.}\label{fig: equality axioms}
\end{figure}

By definition, a \emph{(first-order) semantic specification} of
the object language $\Lang{L}$ is a set $S$ of $\Lang{L}$-open
$\FO(\Lang{L})$-sentences defining the connectives of $\Lang{L}$.
For the sake of generality we always include the standard equality
axioms listed in Figure~\ref{fig: equality axioms} in a semantic
specification~$S$. 
This ensures that $\approx$ is a congruence on every sort in any first-order
interpretation of $\FO(\Lang{L})$.
We assume the set of $\sigma$-definitions with respect to $S$ of all
the connectives $\sigma$ of $\Lang{L}$ is fixed and explicitly given as
the set $S^0$.

Intuitively, a specification $S$ of a semantics of the given logic
is an axiomatisation in the language $\FO(\Lang{L})$
of a class of $\Lang{L}$-structures 
where each $\Lang{L}$-connective $\sigma$ has an unambiguous
representation. 
Because the Beth definability property holds for first-order logic
we can assume that all such representations are explicit, that is,
every connective $\sigma$ is defined by an explicit formula
$\phi^\sigma$.
The collection of explicit definitions of all the connectives
constitutes the set $S^0$.
Since there are many ways of axiomatising the same (axiomatisable)
class of first-order structures and choosing explicit representations
for connectives, there are many ways of specifying a semantics
and choosing a set of semantic definitions for a semantic
specification.
Axiomatisations of the empty class of $\Lang{L}$-structures are all
inconsistent and, hence, semantic specifications can
be inconsistent.

As an example we give a semantic specification for the logic \SO.
Suppose $S_\SO$ consists of the following $\Lang{L}_\SO$-open sentences together
with the default equality axioms.
\smallskip
\begin{trivlist}
 \item[\hskip\parindent] Connective definitions:
\[\setlength{\arraycolsep}{1pt}
\begin{array}{lrcl}%
&\forall x\;\bigl(\nu_1(\{\ell\},x)&\equiv&%
 \nu_0(\ell)\approx x\bigr)\\
&\forall x\;\bigl(\nu_1(\Not p,x)&\equiv&%
 \neg\nu_1(p,x)\bigr)\\
&\forall x\;\bigl(\nu_1(p\Or q,x)&\equiv&%
\nu_1(p,x)\lOr\nu_1(q,x)\bigr)\\
&\forall x\;\bigl(\nu_1(\exists r.p,x)&\equiv&%
\exists y\;\bigl(\nu_2(r,x,y)\lAnd\nu_1(p,y)\bigr)\bigr)
\end{array}
\]
\item[\hskip\parindent] Transitivity axiom:
\[
\forall x\forall y\forall z\;\bigl((\nu_2(r,x,y)\lAnd\nu_2(r,y,z))\lImp\nu_2(r,x,z)\bigr)
\]
\end{trivlist}\smallskip
The first four sentences are the connective definitions of
$\Lang{L}_\SO$ and
constitute the set $S_\SO^0$.
The fifth sentence does not belong to $S_\SO^0$.
It is the transitivity axiom specifying that all role symbols $r$ are
transitive.
If we wanted to specify that only a subset of the role symbols are
transitive, this can be done by including one transitivity axiom
for each role (constant) symbol that is meant to be transitive.

Because, in general, there are many possibilities of axiomatising the same
class of $\Lang{L}$-structures, there are many possibilities for
specifying the semantics of a logic.
In this paper we restrict our attention to semantic specifications
in forms that are standard in the literature for non-classical logics.

We say a semantic specification $S$ is
\emph{normalised},
if it consists of three disjoint parts, that is,
$S=S^+\cup S^-\cup S^b$,
where $S^+$, $S^-$
and $S^b$ are disjoint sets of sentences
satisfying the following:
\begin{nlist}
 \item\label{def: positive connective def} $S^+$ is a set of
$\Lang{L}$-open sentences 
of the form:
\[
\xi^E_+\define
\forall
\seq{x}\;%
(%
\nu_n(E(p_1,\ldots,p_m),\seq{x})\Imp
\phi^E_+(p_1,\ldots,p_m,\seq{x})%
).
\]
\item\label{def: negative connective def} $S^-$ is a set of
$\Lang{L}$-open sentences  
of the form:
\[
\xi^E_-\define
\forall\seq{x}\;%
(%
\phi^E_-(p_1,\ldots,p_m,\seq{x})\Imp
\nu_n(E(p_1,\ldots,p_m),\seq{x})%
).
\]
\item All $\Lang{L}$-expressions occurring in $S^b$ are atomic.
\end{nlist}
Here, $E$ denotes any $\Lang{L}$-expression.

In this definition we assume
that multiple sentences 
of the form~\ref{def: positive connective def} 
for the same
expression~$E$
in $S^+$ and $S^-$
are all equivalently reduced to
a single sentence $\xi^E_+$.
Similarly for~\ref{def: negative connective def} and~$\xi^E_-$.
The intuition is that $S^+$ and $S^-$ define the semantics of the
connectives.
$S^+$~defines it for positive occurrences of
expressions $E$ (with free variables $p_1,\ldots,p_m$), while~$S^-$~
defines it for negative occurrences of expressions~$E$.
We  
refer to~$S^b$ as the \emph{background theory} of the semantics $S$.  
In particular, $S^b$ includes the equality axioms from Figure~\ref{fig: equality axioms}.

A semantic specification in the form $S^0\cup S^b$
can be turned into normalised form
by decomposing each connective definition in $S^0$
into two implications. 
In fact, $S^0$ and \text{$S^+\cup S^-$} play the same role in 
axiomatising $\Lang{L}$-connectives in $\FO(\Lang{L})$
modulo the background theory~$S^b$.

The sample semantic specification $S_{\SO}$ can be normalised
by decomposing the connective definitions in $S_{\SO}^0$ into
$S_\SO^+$-sentences and $S_\SO^-$-sentences as follows. 
\smallskip
\begin{trivlist}
 \item[\hskip\parindent] $S_\SO^+$-sentences:
\[\setlength{\arraycolsep}{1pt}
\begin{array}{lrcl}%
&\forall x\;\bigl(\nu_1(\{\ell\},x)&\lImp&%
 \nu_0(\ell)\approx x\bigr)\\
&\forall x\;\bigl(\nu_1(\Not p,x)&\lImp&%
 \neg\nu_1(p,x)\bigr)\\
&\forall x\;\bigl(\nu_1(p\Or q,x)&\lImp&%
\nu_1(p,x)\lOr\nu_1(q,x)\bigr)\\
&\forall x\;\bigl(\nu_1(\exists r.p,x)&\lImp&%
\exists y\;\bigl(\nu_2(r,x,y)\lAnd\nu_1(p,y)\bigr)\bigr)
\end{array}
\]
\item[\hskip\parindent] $S_\SO^-$-sentences:
\[\setlength{\arraycolsep}{1pt}
\begin{array}{lrcl}%
&\forall x\;\bigl(\nu_0(\ell)\approx x&\lImp&%
 \nu_1(\{\ell\},x)\bigr)\\
&\forall x\;\bigl(\neg\nu_1(p,x)&\lImp&%
 \nu_1(\Not p,x)\bigr)\\
&\forall x\;\bigl(\nu_1(p,x)\lOr\nu_1(q,x)&\lImp&%
\nu_1(p\Or q,x)\bigr)\\
&\forall x\;\bigl(\exists y\;\bigl(\nu_2(r,x,y)\lAnd\nu_1(p,y)\bigr)&\lImp&%
\nu_1(\exists r.p,x)\bigr)\\
\end{array}
\]
\end{trivlist}\smallskip
The background theory $S_{\SO}^b$ of $\SO$ consists of this 
sentence,
\[\forall x\forall y\forall z\;\bigl((\nu_2(r,x,y)\lAnd\nu_2(r,y,z))\lImp\nu_2(r,x,z)\bigr),
\]
specifying transitivity of roles plus the default equality axioms.

It is worth noting that the symbol $E$ in definitions~\ref{def: positive connective def} and~\ref{def: negative connective def} denotes an
arbitrary expression in $\Lang{L}$. This means that $E$ does not
necessarily
have the form $\sigma(p_1,\ldots,p_n)$
where~$\sigma$ is a connective.
For example, a specification might be:
\[
\xi^E_+\define
\forall
x\;
(%
\nu_1(\exists r.\exists r.p,x)\Imp
\nu_1(\exists r.p,x)%
)
\]
In this case $E\define\exists r.\exists r.p$ and
$\phi^E_+\define\nu_1(\exists r.p,x)$.

\smallskip
It is convenient to introduce notation for the set of instantiations of
the right hand sides and left hand sides of the $\xi^E_+$ and $\xi^E_-$,
respectively.
For every $\Lang{L}$-expression $E$, let
\begin{align*}
\Phi^E_+&\define\{\phi^F_+(E_1,\ldots,E_m,\seq{x})\mid%
    \text{$E=F(E_1,\ldots,E_m)$ for some $\xi^{F(p_1,\ldots,p_m)}_+$ from $S$}\} \text{ and}\\
\Phi^E_-&\define\{\phi^F_-(E_1,\ldots,E_m,\seq{x})\mid%
    \text{$E=F(E_1,\ldots,E_m)$ for some $\xi^{F(p_1,\ldots,p_m)}_-$ from $S$}\}.
\end{align*}
Thus, $\Phi^E_+$ (respectively\ $\Phi^E_-$) is the set of instantiations of succedents (respectively\ antecedents) 
of positive (respectively\ negative) specifications in $S$,
where the antecedents (respectively\ succedents) match the given expression $E$.

For example, in the case of our specification for $\SO$ and $E=\exists
r.(p\Or q)$, we have
\[
\Phi^{\exists
r.(p\Or q)}_+=\Phi^{\exists r.(p\Or q)}_-=\{\exists
y\,(\nu_2(r,x,y)\lAnd\nu_1(p\Or q,y))\}.
\]

Let $\prec$ be any ordering
on $\Lang{L}$-expressions.
For any $\Lang{L}$-expression $E$ and any set $X$ of
$\Lang{L}$-expressions we define 
\[
\sub_\prec(E)\define\{E'\mid E'\prec E\} \quad
\text{and} \quad
\sub_\prec(X)\define\bigcup_{E\in X}\sub_\prec(E).
\]
That is, $\sub_\prec(X)$ is the set of all expressions $\prec$-smaller than some expression in $X$.
We often write $\sub_\prec(E_1,\ldots,E_m)$ rather than 
$\sub_\prec(\{E_1,\ldots,E_m\})$.

Any normalised specification $S$ of a semantics
\emph{induces} a relation
$\prec$ on expressions as follows.
Let $\prec$ be the smallest
transitive relation satisfying:
$E'\prec E$ whenever 
$E=F(E_1,\ldots,E_m)$, for some $\Lang{L}$-expressions $E_1,\ldots,E_m$,
and $E'$ occurs in
$\phi^F_+(E_1,\ldots,E_m,\seq{x})$
or $\phi^F_-(E_1,\ldots,E_m,\seq{x})$, respectively, for some
sentence $\xi^{F(p_1,\ldots,p_m)}_+$ or $\xi^{F(p_1,\ldots,p_m)}_-$ in $S$.
The reflexive closure of~$\prec$ is denoted by $\preceq$.

Recall that $S^0$ denotes the set of $\Lang{L}$-open sentences that
define the $\Lang{L}$-connec\-ti\-ves.
A semantic specification $S$ is \emph{well-defined} iff  $S$ is normalised and
the following conditions are all true.
\begin{wdlist}
\item\label{condition: well-defined semantics 1}
$\forall S^0,\forall S^b\models\forall S$,
\item\label{condition: well-defined semantics 2}\label{condition: well-defined semantics: well-founded ordering}
the relation $\prec$ induced by $S$ is a well-founded ordering on $\Lang{L}$-expressions, and
\item\label{condition: well-defined semantics 3}
for every expression $E=\sigma(E_1,\ldots,E_m)$,\\
$
\forall S^0, S^b\range\sub_\prec(E)\models_c
\begin{aligned}[t]
\forall\seq{x}\Bigl(%
&
\Bigl(\bigwedge \Phi^E_+\Imp\phi^\sigma(E_1,\ldots,E_m,\seq{x})\Bigr)\land{}\\%
&
\Bigl(\phi^\sigma(E_1,\ldots,E_m,\seq{x})\Imp\bigvee \Phi^E_-\Bigr)\Bigr).
\end{aligned}
$\\
\end{wdlist}
Condition~\ref{condition: well-defined semantics 3}
follows from the following first-order condition: 
\begin{wdlist}
\item[(wd3$'$)]\label{condition: well-defined semantics 3a}%
for every connective $\sigma$,\\
$%
\forall S^0, S^b\range\sub_\prec(\sigma(\seq{p}))\models_c%
\begin{aligned}[t]
\forall\seq{x}\Bigl(%
&
\Bigl(\bigwedge \Phi^{\sigma(\seq{p}%
)}_+\Imp\phi^\sigma(\seq{p},\seq{x})\Bigr)\land{}\\%
&
\Bigl(\phi^\sigma(\seq{p}%
,\seq{x})\Imp\bigvee \Phi^{\sigma(\seq{p}%
)}_-\Bigr)\Bigr).
\end{aligned}%
$
\end{wdlist}
Because we can assume that $S^0$ is also a normalised semantic specification,
it similarly induces a relation $\prec_0$ that can be assumed to be a well-founded ordering.
Standardly, the semantics of a logic is defined
by induction over the interpretation of the connectives and primitives
(that is, constants, and variables)
which is homomorphically lifted 
to arbitrary $\Lang{L}$-expressions.
This is equivalent to  assuming
a well-founded ordering on expressions of~$\Lang{L}$.
For any reasonable definition of a semantics
such a well-founded ordering exists.
Thus, although it is not difficult to imagine formulae $\phi^\sigma$
such that~$\prec_0$ is not well-founded,
we assume that the~$\phi^\sigma$ are chosen in such a way
that it is possible to lift the semantics of $\Lang{L}$-primitives
to arbitrary $\Lang{L}$-expressions, that is, $\prec_0$ is well-founded.

In the case of $S_{\SO}$, because $S_\SO^+$ and $S_\SO^-$ are obtained 
by decomposing the set $S_\SO^0$,
the two orderings $\prec$ and $\prec_0$
coincide. 
Similar to many cases of description and modal logics,
$\prec$ and $\prec_0$ are both just the direct subexpression ordering on
$\Lang{L}_\SO$. 

There are different semantic specifications
which describe the same class of $\Lang{L}$-structures.
As we have just noted, some semantic specifications
already allow the lifting of the semantics
from atomic expressions to arbitrary $\Lang{L}$-expressions. 
We assume that $S^0\cup S^b$ is such a
specification and implicitly
accommodates $\Lang{L}$-connectives.
According to this definition,
a well-defined semantic specification $S$ is equivalent
to $S^0\cup S^b$ modulo the background theory~$S^b$.
This is ensured by condition~\ref{condition: well-defined semantics 1}
and the assumption that $S$ defines all $\Lang{L}$-connectives
in $S^0$.
Through condition~\ref{condition: well-defined semantics 2}, $S$ imposes its own inductive
structure on $\Lang{L}$-expressions.
Condition~\ref{condition: well-defined semantics 3} specifies a correlation
between $S$ and $S^0$ on instances of $\Lang{L}$-expressions.
It can be seen that $S^0\cup S^b$ is a well-defined semantic specification itself.

Let us consider if the semantic specification of $\SO$ above is well-defined.
The first condition is 
satisfied because
$S_{\SO}=S_{\SO}^0\cup S_{\SO}^b$.
The second condition is satisfied because~$\prec$~is the direct
subexpression ordering.
Condition~\ref{condition: well-defined semantics 3a}
is true for all $\Lang{\SO}$ connectives.
For instance, consider the case of $\sigma=\exists\cdot.\cdot$.
Since $\Phi^{\exists r.p}_+=\Phi^{\exists r.p}_-=\{\exists
y\,(\nu_2(r,x,y)\lAnd\nu_1(p,y))\}$,
the formula
\[
\forall x\,\Bigl(\bigl(\exists y\,(\nu_2(r,x,y)\lAnd\nu_1(p,y))\lImp
\phi^\sigma(r,p,x)\bigr)\lAnd\bigl(\phi^\sigma(r,p,x)\lImp\exists y\,(\nu_2(r,x,y)\lAnd\nu_1(p,y))\bigr)\Bigr),
\]
on the right hand side of condition~\ref{condition: well-defined semantics 3a},
is a tautology. 
In a similar way, the condition~\ref{condition: well-defined semantics 3a}
can be checked for the other connectives.

A \emph{(propositional) logic} $L$ over the language $\Lang{L}$ is a
subset of concepts in $\Lang{L}$ which is closed under arbitrary
substitutions of variables with 
expressions of the
same sorts.
A logic $L$ is \emph{first-order definable} iff there is a
semantic specification~$S_L$ such that $L$ coincides with the
set of all concepts that are valid in all $\Lang{L}$-structures
satisfying~$\forall S_L$, that is, 
\[L=\{C\in\Lang{L}^1\mid \forall
S_L\models_c\forall x\;\nu_1(C,x)\}.\]

For a fixed  semantic specification $S_L$ of a logic $L$, 
if $\I$ is an $\Lang{L}$-structure satisfying $S_L$ then by definition $\I$ is
a \emph{model of $L$} or simply an \emph{$L$-model} (with respect
to~$S_L$).

\section{Synthesising a Tableau Calculus}\label{section_tableau_synthesis}

First, we give the needed basic definitions for the kind of tableau
calculi our method generates.

Let $T$ denote a tableau calculus comprising of a set of inference
rules. 
A \emph{tableau derivation} or \emph{tableau} for $T$ is a finitely branching,
ordered tree whose nodes are sets of formulae in $\FO(\Lang{L})$.
Assuming that~$\mathcal S$ is the input set of concept expressions in $\Lang{L}$
to be tested for satisfiability the root node of the tableau is the
set $\{\nu_1(C,a) \mid C \in \mathcal S\}$, where $a$ denotes a fresh 
constant of the domain sort.
For a finite set $\mathcal{S}$, $a$ can be viewed as the Skolem constant
introduced by Skolemising the $\FO(\Lang{L})$-formula $\exists x\;\bigwedge_{C\in\mathcal{S}}\nu_1(C,x)$.
(This can be naturally expanded to infinite sets of concepts but this is not essential for the paper.)

Successor nodes are constructed in accordance with a set of inference
rules in the calculus.
The inference rules have the general form 
\[
\tableaurule{X_0}{X_1 \vert \ldots \vert X_n},
\]
where both the numerator $X_0$ and all denominators $X_i$ 
are finite sets of negated or unnegated atomic formulae 
in the language $\FO(\Lang{L})$.
The formulae in the numerator are called \emph{premises}, while the
formulae in the denominators are called \emph{conclusions}.
$n$ is called the \emph{branching factor} of the rule.
The numerator and all the denominators are non-empty, but $n$ may be zero, in which case 
the denominators are not present
and the rule is a \emph{closure rule}.
Closure rules are also written $X_0/\bot$.
If the branching factor $n$ is greater than one, the rule is a \emph{branching rule}.
An inference rule is applicable to a selected formula $\phi$ in a
node of the tableau, if $\phi$ together with other
formulae in the node, are simultaneous
instantiations of all the premises of the rule.
Then $n$~successor nodes are created which contain the formulae of
the current node and the appropriate instances of $X_i$.
We assume that \emph{any rule is applied at most once to the same set
of premises}, which is a standard assumption for tableau derivations.

We use the notation $T(\mathcal S)$ for a finished (in the
limit) tableau built
by applying the rules of the calculus $T$ starting with the set
$\mathcal S$ (of $\Lang{L}$-concepts) as input.
That is, we assume that all branches in the tableau are fully expanded and
all applicable rules have been applied in~$T(\mathcal S)$.
We assume that all the rules of the calculus are applied
\emph{non-deterministically to a tableau}.
This means that we do not assume any order of rule application 
and, at any given node, an arbitrary rule is chosen for the node expansion
from all the rules which are applicable to formulae of the node.

In a tableau, a maximal path from the root node is called a
\emph{branch}.
For a branch~$\branch{B}$ of a tableau we write $\phi\in\branch{B}$
to indicate that the formula $\phi$ has
been derived in~$\branch{B}$, that is, $\phi$ belongs to a node of the
branch~$\branch{B}$.
Our notion of a tableau branch can be viewed in two ways.
On the one hand, one can view it as having procedural flavour as
a path of nodes in the tableau derivation.
On the other hand, a branch can
be identified with the set-theoretical union of the nodes in it.

A branch of a tableau is \emph{closed} if a closure rule
has been
applied
in this branch, otherwise the branch is called \emph{open}.
The tableau $T(\mathcal{S})$ is \emph{closed} if all its branches are
closed and $T(\mathcal{S})$ is \emph{open} otherwise.
The calculus $T$ is \emph{sound} iff 
for any (possibly infinite) set of concepts~$\mathcal{S}$, each
$T(\mathcal{S})$ is open whenever $\mathcal{S}$ is satisfiable.
$T$ is \emph{complete} iff for any (possibly infinite) unsatisfiable set of concepts
$\mathcal{S}$ there is a $T(\mathcal{S})$ which is closed.

\bigskip
Now, let $L$ be a first-order definable propositional logic
over $\Lang{L}$ and $S_L$ a well-defined semantic specification of $L$,
that is, conditions~\ref{condition: well-defined semantics 1}--\ref{condition: well-defined semantics 3} hold for $S_L$.
We now describe how tableau rules can be synthesised from $S_L$.
If $S_L$ is not already normalised we first normalise it.
Thus assume $S_L = S^+_L \cup S^-_L \cup S^b_L$.

Next we take a positive specification $\xi^E_+$ in $S_L^+$.
Eliminate existential quantifiers using Sko\-le\-mi\-sa\-tion and equivalently rewrite
$\xi^E_+$ into the following implicational form
\[%
\forall x_1\cdots\forall x_n\;%
\left(
\nu_n(E(p_1,\ldots,p_m),x_1,\ldots,x_n)\rightarrow
\bigvee_{j=1}^{J}\bigwedge_{k=1}^{K_j} \psi_{jk}
\right),
\]
where
each $\psi_{jk}$ denotes a literal.
This is always possible.
The implication is now turned into the rule:
\[
\rho_+(\xi^E_+)\define
\tableaurule{\nu_n(E(p_1,\ldots,p_m),x_1,\ldots,x_n)\tand y_1\approx y_1\tand\ldots\tand y_s\approx y_s
}%
{\psi_{11}\tand\ldots\tand\psi_{1K_1}\tor\cdots\tor\psi_{J1}\tand\ldots\tand\psi_{JK_J}},
\]
where $y_1,\ldots,y_s$ denote the free variables occurring in $\psi_{jk}$
which do not occur among the variables $x_1,\ldots,x_n$.
Essentially, the antecedent of the implication has become the main
premise in the numerator
and the succedent has been turned into the
denominators of the rule. 
We say the rule \emph{corresponds} to $\xi^E_+$.
This is repeated for each positive specification in~$S_L^+$.

Analogously, we generate a tableau rule for each negative
specification~$\xi^E_-$ in~$S_L^-$.
The corresponding rules have the form
\[
\rho_-(\xi^E_-)\define
\tableaurule{\Not\nu_n(E(p_1,\ldots,p_m),x_1,\ldots,x_n)\tand y_1\approx y_1\tand\ldots\tand y_s\approx y_s
}%
{\psi_{11}\tand\ldots\tand\psi_{1K_1}\tor\cdots\tor\psi_{J1}\tand\ldots\tand\psi_{JK_J}}.
\]
This is obtained by Skolemising the contrapositive of $\xi^E_-$ 
and then equivalently rewriting it to an implication of the form
\[%
\forall x_1\cdots\forall x_n\;%
\left(
\Not\nu_n(E(p_1,\ldots,p_m),x_1,\ldots,x_n)\rightarrow
\bigvee_{j=1}^{J}\bigwedge_{k=1}^{K_j} \psi_{jk}
\right),
\]
where
each 
$\psi_{jk}$ denotes a literal.

We refer to the rules $\rho_+(\xi^E_+)$ and $\rho_-(\xi^E_-)$ generated
in this way, as the
\emph{decomposition rules}.

If the right hand sides of the implicational forms contain free
variables $y_i$ then these are assumed to be universally quantified
and the generated rules are $\gamma$-rules in the Smullyan classification.
Our use of the equalities $y_i\approx y_i$ in the premises of the
generated rules is a bit non-standard, and can be omitted if this
is preferred.
We use the equalities to achieve domain predication, which makes
explicit that applying $\gamma$-rules only instantiates with terms
(domain elements) that occur on the current branch.

The sentences in the background theory of $S_L$
are turned into rules by first
equivalently transforming them into Skolemised disjunctive normal
form.
More specifically, let $\xi$ be an arbitrary sentence in $S_L^b$.
It is first equivalently rewritten to
\begin{gather}
\label{theory_dnf}
\forall x_1\cdots\forall x_n
\bigvee_{j=1}^{J}\bigwedge_{k=1}^{K_j}\psi_{jk}(p_1,\ldots,p_m,x_1,\ldots,x_n),%
\end{gather}
where each $\psi_{jk}$ denotes a literal, and is then turned into the
corresponding rule, namely
\[
\rho(\xi)\define\tableaurule{p_1\approx p_1\tand\ldots\tand p_m\approx p_m\tand x_1\approx x_1\tand\ldots\tand x_n\approx x_n}%
{\psi_{11}\tand\ldots\tand\psi_{1K_1}\tor\cdots\tor\psi_{J1}\tand\ldots\tand\psi_{JK_J}}.
\]
The $p_1,\ldots,p_m,x_1,\ldots,x_n$ are the variables appearing in~(\ref{theory_dnf}).
The purpose of the equalities in the premises is domain
predication and can optionally be omitted.
Rules corresponding to sentences in $S^b_L$ are called \emph{theory rules}.

For example,\label{example: generated E-rules}
the generated decomposition rules for the existential
restriction operator in the description logic \SO  
are
\begin{gather*}
\tableaurule{\nu_1(\exists r.p,x)}{\nu_2(r,x,f(r,p,x))\tand\nu_1(p,f(r,p,x))}
\quad\text{and}\quad
\tableaurule{\Not\nu_1(\exists r.p,x)\tand y\approx y}{\Not\nu_2(r,x,y)\tor\Not\nu_1(p,y)}.
\end{gather*}
$f(r,p,x)$ in the left rule is the Skolem term introduced for the
quantifier $\exists y$ in the connective definition of $\exists\cdot.\cdot$.
The intuition is that for each $r$, each $p$ and each $x$ matching the
premise of the rule there is an
element $f(r,p,x)$ so that the conclusions of the rule are both true.
The transitivity property for roles in the background theory of the
semantic specification of \SO is transformed to the rule
   \[
    \tableaurule{r\approx r\tand x\approx x\tand y\approx y\tand z\approx z}%
      {\lNot \nu_2(r,x,y)\tor \lNot \nu_2(r,y,z) \tor \nu_2(r,x,z)}.
   \]
These rules are not the familiar rules used in standard description logic
tableau systems, but
in Section~\ref{section_optimisation} we see how to get those by rule refinement.

\begin{figure}[tb]%
\begin{gather*}
\begin{aligned}
& \begin{gathered}
  \tableaurule{\vphantom{Pp}P(x_1,\ldots,x_n)}{\vphantom{Pp}x_1\approx x_1\tand\ldots\tand x_n\approx x_n}\\
  \tableaurule{\vphantom{Pp}\nu_n(p,x_1,\ldots,x_n)}{\vphantom{Pp}p\approx p\tand x_1\approx x_1\tand\ldots\tand x_n\approx x_n}\\
  \tableaurule{\vphantom{Pp}x\approx y}{\vphantom{Pp}y\approx x}\\
  \tableaurule{\vphantom{Pp}P(x_1,\ldots,x_n)\tand x_i\approx y_i}{\vphantom{Pp}P(x_1,\ldots,x_{i-1},y_i,x_{i+1},\ldots,x_n)}
 \end{gathered}
&\qquad \begin{gathered}
  \tableaurule{\vphantom{Pp}\Not P(x_1,\ldots,x_n)}{\vphantom{Pp}x_1\approx x_1\tand\ldots\tand x_n\approx x_n}\\
  \tableaurule{\vphantom{Pp}\Not\nu_n(p,x_1,\ldots,x_n)}{\vphantom{Pp}p\approx p\tand x_1\approx x_1\tand\ldots\tand x_n\approx x_n}\\
  \tableaurule{\vphantom{Pp}x\approx y\tand y\approx z}{\vphantom{Pp}x\approx z}\\
  \tableaurule{\vphantom{Pp}\nu_n(p,x_1,\ldots,x_n)\tand x_i\approx y_i}{\vphantom{Pp}\nu_n(p,x_1,\ldots,x_{i-1},y_i,x_{i+1},\ldots,x_n)}
 \end{gathered}
\end{aligned}\\
\tableaurule{\vphantom{Pp}f(p_1,\ldots,p_m,x_1,\ldots,x_n)\approx f(p_1,\ldots,p_m,x_1,\ldots,x_n)\tand%
             x_i\approx y_i%
           }%
           {\vphantom{Pp}f(p_1,\ldots,p_m,x_1,\ldots,x_n)\approx f(p_1,\ldots,p_m,x_1,\ldots x_{i-1},y_i,x_{i+1},\ldots,x_n)}
\end{gather*}
\caption{Default equality rules for predicates and functions occurring in $S_L$.}\label{fig: equality rules}
\end{figure}

The \emph{equality rules} are generated in essentially the same way from
the equality axioms in the background theory and are refined
in accordance with the method described in Section~\ref{section_optimisation}.
Figure~\ref{fig: equality rules} lists the full set of the \emph{refined} equality
rules included by default in the generated tableau calculus.

Since in our formalisation the equality predicate(s) are also used as domain
predicate(s) in order to keep track of the ground terms that occur in
the tableau branches, we include rules which ensure that expressions of the form $t\approx t$
are treated as domain predicates and appear in every branch
of a tableau for every term $t$ in the branch.
These are the first four rules in Figure~\ref{fig: equality rules}.
In particular, these rules ensure that for any term occurring in a
literal $(\neg) P(t_1,\ldots,t_n)$ or $(\neg) \nu_n(q,t_1,\ldots,t_n)$
on any branch, the equalities $t_1\approx t_1\tand\ldots\tand
t_n\approx t_n$ and $q\approx q$ are added to the branch.
The rules also state reflexivity of the equality predicate(s).
The remaining rules are variations of standard rules for equality.
The rules in row three and four ensure that $\approx$ is a congruence
relation for predicates on terms occurring in a branch.
The rule in the last row is a congruence rule for function symbols $f$ occurring
in a branch including Skolem function symbols.

We note that the equality predicate $\approx$ is treated as an ordinary
constant predicate symbol of the meta-language $\FO(\Lang{L})$ and,
hence, can occur in any place where an ordinary predicate symbol $P$ can
occur.

Finally the generated tableau calculus also includes the following
\emph{closure rules}.
\begin{align}
\label{closure_rules}
 \tableaurule{\nu_n(p,\seq{x}%
    )\tand\Not\nu_n(p,\seq{x}%
    )}{\bot}\qquad\qquad
 \tableaurule{P(\seq{x}%
    )\tand\Not P(\seq{x}%
    )}{\bot}
\end{align}
for each sort $n$ and every constant predicate symbol $P$ occurring
in the semantic specification~$S_L$ of the logic.

We use $T_L$ to denote the generated tableau calculus.
In summary, it  
consists of these rules.
\begin{tlist}
\item
The decomposition rules $\rho_+^\sigma(\xi)$ and $\rho_-^\sigma(\xi')$
corresponding to all positive specifications~$\xi$ in~$S_L^+$ and all
negative specifications $\xi'$ in $S_L^-$.
\item
The theory rules $\rho(\zeta)$
corresponding to all sentences $\zeta$
in the background theory~$S_L^b$. 
\item
The equality rules of Figure~\ref{fig: equality rules}.
\item
The closure rules (\ref{closure_rules}).
\end{tlist}\medskip

\noindent Note for each connective there are exactly two decomposition rules in
the calculus $T_L$, one for unnegated occurrences and one for negated
occurrences of the connective.

\begin{figure}
\begin{trivlist}
\item
Decomposition rules:
\begin{gather*}
\begin{alignedat}{1}
&\tableaurule{\nu_1(\{\ell\},x)}{\nu_0(\ell)\approx x}%
&\qquad\quad
&\tableaurule{\Not\nu_1(\{\ell\},x)}{\nu_0(\ell)\not\approx x}%
&\qquad\quad
&\tableaurule{\nu_1(\Not p,x)}{\Not\nu_1(p,x)}%
&\qquad\quad
&\tableaurule{\Not\nu_1(\Not p,x)}{\nu_1(p,x)}%
\end{alignedat}
\\
\begin{alignedat}{1}
&\tableaurule{\nu_1(p_1\lOr p_2,x)}{\nu_1(p_1,x)\tor\nu_1(p_2,x)}
&\qquad\quad%
&\tableaurule{\Not\nu_1(p_1\lOr p_2,x)}{\Not\nu_1(p_1,x)\tand\Not\nu_1(p_2,x)}
\end{alignedat}
\\
\begin{alignedat}{1}
\tableaurule{\nu_1(\exists r.p,x)}{\nu_2(r,x,f(r,p,x))\tand\nu_1(p,f(r,p,x))}
&\qquad\quad
\tableaurule{\Not\nu_1(\exists r.p,x)\tand y\approx y}%
                {\neg \nu_2(r,x,y)\tor\neg\nu_1(p,y)}
\end{alignedat}
\end{gather*}

\item
Transitivity rule:
\begin{gather*}
\tableaurule{r\approx r\tand x\approx x\tand y\approx y\tand z\approx z}%
    {\lNot \nu_2(r,x,y)\tor \lNot \nu_2(r,y,z) \tor \nu_2(r,x,z)}
\end{gather*}
\item
Equality congruence rules:
\begin{gather*}
  \tableaurule{\vphantom{Pp}x\approx y}{\vphantom{Pp}x\approx x\tand y\approx y}
  \qquad\quad
  \tableaurule{\vphantom{Pp}x\not\approx y}{\vphantom{Pp}x\approx x\tand y\approx y}
  \qquad\quad
  \tableaurule{\vphantom{Pp}x\approx y}{\vphantom{Pp}y\approx x}
  \qquad\quad
  \tableaurule{\vphantom{Pp}x\approx y\tand y\approx z}{\vphantom{Pp}x\approx z}\\
  \tableaurule{\vphantom{Pp}\nu_1(p,x)}{\vphantom{Pp}p\approx p\tand x\approx x}
  \qquad
  \tableaurule{\vphantom{Pp}\Not\nu_1(p,x)}{\vphantom{Pp}p\approx p\tand x\approx x}
  \qquad
  \tableaurule{\vphantom{Pp}\nu_2(r,x,y)}{\vphantom{Pp}r\approx r\tand x\approx x\tand y\approx y}
  \qquad 
  \tableaurule{\vphantom{Pp}\Not\nu_2(r,x,y)}{\vphantom{Pp}r\approx r\tand x\approx x\tand y\approx y}\\%
  \tableaurule{\vphantom{Pp}\nu_1(p,x)\tand x\approx y}{\vphantom{Pp}\nu_1(p,y)}
  \qquad\quad
  \tableaurule{\vphantom{Pp}\nu_2(r,x,y)\tand x\approx z}{\vphantom{Pp}\nu_2(r,z,y)}
  \qquad\quad
  \tableaurule{\vphantom{Pp}\nu_2(r,x,y)\tand y\approx z}{\vphantom{Pp}\nu_2(r,x,z)}\\
  \tableaurule{\vphantom{Pp} f(r,p,x)\approx f(r,p,x)\tand x\approx y}%
           {\vphantom{Pp}f(r,p,x)\approx f(r,p,y)}
\end{gather*}
\item
Closure rules:
\begin{gather*}
\begin{alignedat}{1}
\tableaurule{\nu_1(p,x)\tand\Not\nu_1(p,x)}{\bot}%
&\qquad\quad
\tableaurule{\nu_2(r,x,y)\tand\Not\nu_2(r,x,y)}{\bot}%
&\qquad\quad
\tableaurule{x\approx y\tand x\not\approx y}{\bot}%
\end{alignedat}
\end{gather*}
\end{trivlist}
\caption{Generated tableau rules for \SO.}\label{fig: SO tableau}
\end{figure}

For \SO the described approach generates the tableau rules listed in
Figure~\ref{fig: SO tableau}.

\section{Ensuring Soundness and Constructive Completeness}
\label{section_sound_constructively_complete}

We first prove soundness of the calculus $T_L$ synthesised 
from a normalised semantic specification $S_L$.
It is possible to prove that every rule of the generated
calculus~$T_L$ preserves satisfiability of $\FO(\Lang{L})$-formulae.
That is, if all premises of a rule 
are true in an $L$-model $\I$ (under a canonical valuation)
then the conclusions of some branch are also true.
This is the case because the transformation of the semantic specification ensures that the definitions of the rules
basically mimic the semantic definitions.
Hence, soundness is ensured. 

\begin{theorem}[Soundness]\label{theorem_soundness}
Let $T_L$ be a tableau calculus generated from a
normalised semantic specification $S_L$ of a logic $L$.
Then $T_L$ is sound for $L$, that is, 
for every possibly infinite set of concepts $\mathcal{S}$ satisfiable in an $L$-model, 
any finished tableau derivation $T_L(\mathcal{S})$ is open.
\end{theorem}

Now, we prove constructive completeness of $T_L$.
Let $\branch{B}$ denote an arbitrary branch in a $T_L$-tableau
derivation.
We define the following relation $\simB$ with respect to $\branch{B}$:
\[t\simB t'\defiff t\approx t'\in\branch{B},\]
for any ground terms $t$ and $t'$ of the domain sort
$N+1$ in~$\branch{B}$.
Let $\ecl{t}\define\{t'\mid t\simB t'\}$ be the equivalence class of an element $t$.
The presence of the rules of Figure~\ref{fig: equality rules}
ensures that~$\simB$ is a congruence
relation on all domain ground terms in~$\branch{B}$.

We say a model~$\I$, under  a (canonical) valuation~$\iota$,
\emph{reflects} an expression $E$ of the sort~$n$
occurring in
a branch $\branch{B}$
iff
for all
ground terms $t_1,\ldots,t_n$ we have that
\begin{enumerate}[$\bullet$]
       \item
           $(E,\iota(t_1),\ldots,\iota(t_n))\in\nu_n^\I$ whenever
           $\nu_n(E,t_1,\ldots,t_n)\in\branch{B}$, and
       \item
           $(E,\iota(t_1),\ldots,\iota(t_n))\notin\nu_n^\I$ whenever $\Not\nu_n(E,t_1,\ldots,t_n)\in\branch{B}$.
\end{enumerate}\medskip

\noindent Similarly, $\I$ \emph{reflects} predicate constant $P$ from $\branch{B}$
under a (canonical) valuation~$\iota$ in $\I$
iff
for all
ground terms $t_1,\ldots,t_n$ we have that
\begin{enumerate}[$\bullet$]
 \item $(\iota(t_1),\ldots,\iota(t_n))\in P^\I$ whenever $P(t_1,\ldots,t_n)\in\branch{B}$,
            and 
 \item $(\iota(t_1),\ldots,\iota(t_n))\notin P^\I$ whenever $\Not P(t_1,\ldots,t_n)\in\branch{B}$.
\end{enumerate}\medskip

\noindent A model $\I$ \emph{reflects} branch $\branch{B}$ under a valuation $\iota$,
if $\I$ reflects all predicate constants and expressions occurring in $\branch{B}$
under $\iota$.

A tableau calculus $T_L$ is said to be \emph{constructively complete} (for $L$)
iff for any given set of concept $\mathcal{S}$, 
if $\branch{B}$ is an open branch in a tableau derivation~$T_L(\mathcal{S})$
then there is an $L$-model $\I$ 
such that: 
\begin{mlist}
     \item The domain $\Delta^{\I}$ of $\I$ is the set of the equivalence classes
           $\ecl{t}$ for each ground term $t$ occurring in $\branch{B}$.
     \item $\I$ reflects $\branch{B}$ under the \emph{canonical projection valuation} $\pi$ defined by
            $\pi(t)\define\ecl{t}$, for every ground term $t$ occurring in $\branch{B}$.
\end{mlist}
It is clear that if $T_L$ is constructively complete then $T_L$ is complete for $L$.

Suppose now that $S_L$ is a
well-defined  
semantic specification and $\prec_0$ is a well-founded ordering on
$\Lang{L}$-expressions induced by the set $S_L^0$
of the definitions of the connectives
of the form~\eqref{eq: connective def} with respect to $S_L$.

Let $\branch{B}$ be an open branch in a finished tableau derivation in
$T_L$.
We define interpretations of predicate symbols in $\IB$
by induction on $\prec_0$ as follows: 
\begin{enumerate}[$\bullet$]
 \item 
For every $n$-ary constant predicate symbol $P$ in $S_L$,
        \[P^\IB\define\{(\ecl{t_1},\ldots,\ecl{t_n})\mid P(t_1,\ldots,t_n)\in\branch{B}\}.\]
 \item For every $n=1,\ldots,N$ the interpretation $\nu_n^\IB$ of 
       the $\nu_n$ symbols is defined as the smallest subset of
       $\Lang{L}^n\times(\Delta^\IB)^n$
       satisfying both the following, 
       for every variable or constant $p$ of the sort $n$, 
       every connective $\sigma$, and any expressions $E_1,\ldots,E_m$:
\begin{align*}
       (p,\ecl{t_1},\ldots,\ecl{t_n})\in\nu_n^\IB & \iff%
                    \nu_n(p,t_1,\ldots,t_n)\in\branch{B},
\\
       \makebox[11.2em][l]{$(\sigma(E_1,\ldots,E_m),\ecl{t_1},\ldots,\ecl{t_n})\in\nu_n^\IB$} \\
       & \iff \IB\models_c\phi^\sigma(E_1,\ldots,E_m,\ecl{t_1},\ldots,\ecl{t_n}).
\end{align*}
\end{enumerate}\medskip

\noindent In what follows, 
we say that $\IB$ reflects an expression $E$ (a predicate $P$, or a branch $\branch{B}$)
if $\IB$ reflects $E$ ($P$, or $\branch{B}$, respectively)
under the canonical projection valuation $\pi$,
and omit any explicit reference to $\pi$.

A consequence of the definition of $\IB$ is that the definitions of
the connectives are valid in $\IB$: 
\begin{lemma}
$\IB\models\forall S^0_L$.
\end{lemma}

\begin{lemma}\label{lemma: SLb induction}
Let $X$ be any set of expressions occurring in $\branch{B}$.
Suppose $\IB$ reflects all the expressions from $X$.
Then
$\IB\models_c S_L^b\range X$.
\end{lemma}
\begin{proof}
Consider any $\xi\in S_L^b$ and suppose the Skolemised form of $\xi$ is as in~\eqref{theory_dnf}, that is:
\[\xi(p_1,\ldots,p_m)\equiv%
\forall x_1\cdots\forall x_n\;\bigvee_{j=1}^{J}\bigwedge_{k=1}^{K_j} \psi_{jk}(p_1,\ldots,p_m,x_1,\ldots,x_n).\]
Let $E_1,\ldots,E_m$ be any expressions from $X$ 
and $t_1,\ldots,t_n$ any ground terms of sort $N+1$
occurring in $\branch{B}$.
By rule $\rho(\xi)$,
there is a $j=1,\ldots,J$
such that, for all $k=1,\ldots,K_j$, the literals $\psi_{jk}(E_1,\ldots,E_m,t_1,\ldots,t_n)$
are in the branch~$\branch{B}$.
Since~$S_L^b$ does not contain
non-atomic
expressions of the language $\Lang{L}$
we have that $\IB\models_c\psi_{jk}(E_1,\ldots,E_m,\|t_1\|,\ldots,\|t_n\|)$ 
by the assumptions of the lemma for every $k=1,\ldots,K_j$.
This implies that $\IB\models_c\xi(E_1,\ldots,E_m,\|t_1\|,\ldots,\|t_n\|)$.
\end{proof}

\begin{corollary}\label{corr: SLb}
$\IB\models_c S_L^b$. 
\end{corollary}
\begin{proof}
From the definition of $\IB$ 
and the closure rules
we get that 
$P(t_1,\ldots,t_n)\in\branch{B}$ implies $(\ecl{t_1},\ldots,\ecl{t_n})\in P^\IB$,
$\Not P(t_1,\ldots,t_n)\in\branch{B}$ implies $(\ecl{t_1},\ldots,\ecl{t_n})\notin P^\IB$,
$\nu_n(p,t_1,\ldots,t_n)\in\branch{B}$ implies $(p,\ecl{t_1},\ldots,\ecl{t_n})\in \nu_n^\IB$, and 
$\Not \nu_n(p,t_1,\ldots,t_n)\in\branch{B}$ implies $(p,\ecl{t_1},\ldots,\ecl{t_n})\notin \nu_n^\IB$
for every constant predicate symbol $P$, $n=0,\ldots,N$, and primitive $p$
of sort $n$. Thus, $\IB\models_c S_L^b$ by Lemma~\ref{lemma: SLb induction}.
\end{proof}

\begin{lemma}\label{lemma: IB reflects B}
$\IB$ reflects the branch $\branch{B}$. 
\end{lemma}
\begin{proof}

By simultaneous induction on the well-founded ordering $\prec$ induced by $S_L$ 
we show that for all $n=1,\ldots,N$, for every $E$,
and all domain ground terms $t_1,\ldots,t_n$
(of sort $N+1$) in $\branch{B}$, we have that
\begin{enumerate}[$\bullet$]
 \item $(E,\ecl{t_1},\ldots,\ecl{t_n})\in\nu_n^\IB$ whenever $\nu_n(E,t_1,\ldots,t_n)\in\branch{B}$, and
 \item $(E,\ecl{t_1},\ldots,\ecl{t_n})\notin\nu_n^\IB$ whenever $\Not\nu_n(E,t_1,\ldots,t_n)\in\branch{B}$.
\end{enumerate}\medskip

\noindent We have the following two cases which correspond to the base case of the induction and to the induction step:

Case $E=p$. This case follows from the definition of $\IB$.

Case $E=\sigma(E_1,\ldots,E_m)$.
        Suppose $\nu_n(E,t_1,\ldots,t_n)\in\branch{B}$.
        Let $\xi^F_+$ be such that $E=F(F_1,\ldots,F_m)$ for some $F_1,\ldots,F_m$
               and the Skolemised form of the
               corresponding $\phi^F_+$ is as follows
\[%
\phi^F_+(p_1,\ldots,p_m,x_1,\ldots,x_n)\equiv%
\bigvee_{j=1}^{J}\bigwedge_{k=1}^{K_j} \psi_{jk}(p_1,\ldots,p_m,x_1,\ldots,x_n).
\]
Then by rule $\rho_+(\xi^F_+)$
there is a $j=1,\ldots,J$ such that, for all $k=1,\ldots,K_j$, the literals $\psi_{jk}(F_1,\ldots,F_m,t_1,\ldots,t_n)$
are in $\branch{B}$.
Further, for every expression $E'(F_1,\ldots,F_m)$ which
occurs in $\psi_{jk}(F_1,\ldots,F_m,t_1,\ldots,t_n)$, where $k=1,\ldots,K_j$,
we have $E'(F_1,\ldots,F_m)\prec F(F_1,\ldots,F_m)=E$.
Thus, by the induction hypothesis, for every $k=1,\ldots,K_j$, $\IB\models_c\psi_{jk}(F_1,\ldots,F_m,\ecl{t_1},\ldots,\ecl{t_n})$.
Consequently, we have \[\IB\models_c\phi^F_+(F_1,\ldots,F_m,\ecl{t_1},\ldots,\ecl{t_n})\]
and, hence, $\IB\models_c\Phi^E_+(\ecl{t_1},\ldots,\ecl{t_n})$.
By Lemma~\ref{lemma: SLb induction},
$\IB\models_c S_L^b\range\sub_\prec(E)$.
Since $\IB\models\forall S_L^0$, 
we obtain
$\IB\models_c\phi^\sigma(E_1,\ldots,E_m,\ecl{t_1},\ldots,\ecl{t_n})$
and, therefore, by the definition of $\IB$, we have $(E,\ecl{t_1},\ldots,\ecl{t_n})\in\nu_n^\IB$.

The second implication for negative literals is proved similarly.
\end{proof}

As a consequence we obtain the following theorem.
\begin{theorem}[Constructive completeness]
\label{theorem_constructive_completeness}
Let $T_L$ be a tableau calculus generated from a
well-defined semantic specification $S_L$ of a logic $L$.
Then $T_L$ is constructively complete. 
\end{theorem}
\begin{proof}
We only need to prove that $\IB\models\forall S_L$.
However, this follows from $\forall S^0_L,\forall S^b_L\models\forall S_L$
since $\IB\models\forall S^b_L$ by Lemma~\ref{lemma: IB reflects B} and Lemma~\ref{lemma: SLb induction}.
\end{proof}

\section{Refining Rules by Turning Conclusions into Premises}
\label{section_optimisation}

Generally the degree of branching of the generated rules is higher than is
necessary.
Furthermore, representation of the generated rules
involves the additional symbols of the language $\FO(\Lang{L})$
creating a syntactic overhead which may not always be justified.
To address these problems
in this section, and the next, we introduce two techniques for refining
the generated rules.

The first technique reduces the number of branches of a rule by
constraining the rule with additional premises and deriving fewer 
conclusions.
Suppose~$r$ is a tableau rule in a sound and constructively complete
tableau calculus $T_L$.
Suppose $r$ has this form.
\[
 r\define\tableaurule{X_0}{X_1\tor\cdots\tor X_m}.
\]
Let $X_i=\{\psi_1,\ldots,\psi_k\}$ be one of the denominators of the
rule $r$ for some $i\in\{1,\ldots,m\}$.
Without loss of generality we assume that $i=1$.

Consider the rules $r_j$ with $j=1,\ldots,k$ defined by
\[
 r_j\define\tableaurule{X_0\cup\{\inverse\psi_j\}}{X_2\tor\cdots\tor X_m}.
\]
Each $r_j$ is obtained from the rule $r$ by removing the first
denominator $X_1$ and adding the negation of one of the formulae in $X_1$ as a premise.
We can drop any domain predication equalities from the numerator when they are not necessary.

Let $r$ denote a rule in $T_L$.
We denote by $\Rfr(r,T_L)$ the \emph{refined tableau calculus} obtained from $T_L$
by replacing the rule $r$ with rules $r_1,\ldots,r_k$. 
It is clear that the calculus $\Rfr(r,T_L)$ is sound. 
In general, $\Rfr(r,T_L)$ is however not constructively complete.
Nevertheless, analysis of the proofs 
of Lemma~\ref{lemma: IB reflects B} and Lemma~\ref{lemma: SLb induction}
shows that the following theorem is true.

\begin{theorem}\label{theorem: tableau transformation}
Let $T_L$ be a tableau calculus generated from a
well-defined specification $S_L$ of the logic $L$.
Let $r$ be the rule ${X_0}/{X_1\tor\cdots\tor X_m}$ in $T_L$ and suppose
$\Rfr(r,T_L)$ is a refined version of $T_L$. 
Further, suppose $\branch{B}$ is an open branch in a $\Rfr(r,T_L)$-tableau
derivation
and 
 for every set $Y$ of $\Lang{L}$-expressions from $\branch{B}$ the following holds.
\smallskip
\begin{trivlist}
 \item[\hskip\parindent]
    If all expressions in $Y$ are reflected in $\IB$
    then for every $E_1,\ldots,E_l\in Y$,
\begin{align*}
    \begin{aligned}
        & X_0(E_1,\ldots,E_l,t_1,\ldots,t_n)\subseteq\branch{B}\ \text{implies}\ \\
        & \IB\models X_i(E_1,\ldots,E_l,\ecl{t_1},\ldots,\ecl{t_n}),\ \text{for some}\ i=1,\ldots,m.
    \end{aligned}
\tag{$\dagger$}\label{condition: tableau transformation}
\end{align*}
\end{trivlist}\smallskip
Then, $\branch{B}$ is reflected in $\IB$.
\end{theorem}

Roughly, condition~\eqref{condition: tableau transformation} says that
the replaced rule $r$ is admissible in the model $\IB$ associated
with $\branch{B}$ constructed using the refined calculus $\Rfr(r,T_L)$. 
An immediate consequence is the following.

\begin{corollary}\label{corollary: refinement preserves constructive completeness}
If the condition of Theorem~\ref{theorem: tableau transformation}
holds for every open branch $\branch{B}$ of any $\Rfr(r,T_L)$-tableau
derivation
then the refined calculus $\Rfr(r,T_L)$ is constructively complete.
\end{corollary}
Generalising this refinement to turning more than one denominator into
premises is not difficult. 
Theorem~\ref{theorem: tableau transformation} 
can be reformulated to accommodate this generalisation
and the formulation of Corollary~\ref{corollary: refinement preserves constructive completeness}
does not change then.

We observe that the condition~\eqref{condition: tableau transformation} is implied by 
the following condition:
\begin{gather}
    \begin{aligned}
        &\text{if}\ X_0(E_1,\ldots,E_l,t_1,\ldots,t_n)\subseteq\branch{B}\ \text{and}\ 
        \IB\not\models X_1(E_1,\ldots,E_l,\ecl{t_1},\ldots,\ecl{t_n})\\ 
        &\text{then}\ 
        X_i(E_1,\ldots,E_l,t_1,\ldots,t_n)\subseteq\branch{B},\ \text{for some}\ i=2,\ldots,m.
    \end{aligned}%
\tag{$\ddagger$}\label{condition: tableau transformation1}
\end{gather}
This follows by an induction argument on the well-founded ordering $\prec$.

For example, consider the generated rule for negative occurrences of the existential
restriction operator given in
Section~\ref{section_tableau_synthesis}. 
\[
\tableaurule{\Not\nu_1(\exists r.p,x)\tand y\approx y}{\Not\nu_2(r,x,y)\tor\Not\nu_1(p,y)}
\]
In most description logics it can be replaced with the more often
seen rule:
\begin{align*}
\tableaurule{\Not\nu_1(\exists r.p,x)\tand\nu_2(r,x,y)}{\Not\nu_1(p,y)}.
\end{align*}
In such cases, 
condition~\eqref{condition: tableau transformation1} 
has the following form.
    \begin{equation*}
        \text{If}\ \Not\nu_1(\exists E.F,t)\in\branch{B}\ \text{and}\ 
        \IB\models\nu_2(E,t,t')\ 
        \text{then}\ 
        \Not\nu_1(F,t')\in\branch{B}.
    \end{equation*} 
For description and modal logics such as \SO
the proof of this condition is typically part of 
the proof of the completeness theorem for the calculus
which is standardly proved by induction on the well-founded relation $\prec$
(or equivalently, by induction on the way formulae are derived on a branch).
For \SO condition~\eqref{condition: tableau transformation1} can be proved separately
and implies that condition~\eqref{condition: tableau transformation}
is true for every branch of the refined tableau.
Thus, this rule refinement preserves constructive completeness.

The default equality rules (given in Figure~\ref{fig: equality rules})
added to every generated calculus are already in refined form.
The rules that would be produced from the semantic specification of equality
in Figure~\ref{fig: equality axioms} have a different form.
For example, 
the congruence rule
\[
\tableaurule{\nu_n(p,\seq{x})\tand x_i\approx y_i}{\nu_n(p,x_1,\ldots,x_{i-1},y_i,x_{i+1},\ldots,x_n)}
\]
is a refined form (obtained in two steps) of the following rule:
\[
\tableaurule{p\approx p\tand x_1\approx x_1\tand\ldots\tand x_n\approx x_n\tand
y_i\approx y_i}{\Not\nu_n(p,\seq{x})\tor x_i\not\approx
y_i\tor\nu_n(p,x_1,\ldots,x_{i-1},y_i,x_{i+1},\ldots,x_n)}.
\]

Transitivity of a role $r$ provides another example where
rule refinement converts the rule
\[
    \tableaurule{r\approx r\tand x\approx x\tand y\approx y\tand z\approx z}%
      {\lNot \nu_2(r,x,y)\tor \lNot \nu_2(r,y,z) \tor \nu_2(r,x,z)}
\]
into the more familiar rule 
\[
\tableaurule{\nu_2(r,x,y)\tand\nu_2(r,y,z)}{\nu_2(r,x,z)}.
\]
Condition~\eqref{condition: tableau transformation} holds in this case
since it follows from the definition of $\IB$ that $\IB$ reflects all
atomic formulae of the form $\nu_2(r,x,y)$ for any role symbol $r$
in the branch $\branch{B}$.

As a negative example let us consider the possibility of replacing the rule for disjunction
\[
 \tableaurule{\nu_1(p\Or q, x)}{\nu_1(p,x)\tor\nu_1(q,x)}[$\Or$]\label{rule: or}
\]
by this rule.
\[
 \tableaurule{\nu_1(p\Or q, x)\tand\Not \nu_1(p,x)}{\nu_1(q,x)}[$\Or'$]\label{rule: KE or}
\]
In KE tableau calculi this rule is used together with an analytic
cut rule~\cite{DAgostinoMondadori-TC+-1994}.
This raises the question whether a cut rule is essential for
completeness and whether the~\eqref{rule: KE or}-rule alone would
suffice instead of~\eqref{rule: or}.

Consider a tableau calculus $T$ without any other rules to decompose
positive occurrences of disjunctions except the standard rule~\eqref{rule: or}.
Suppose $T'$ is the calculus where the \eqref{rule: or}-rule has been
replaced by the \eqref{rule: KE or}-rule. That is, $T'\define\Rfr(\eqref{rule: or},T)$.
Examination reveals that condition~\eqref{condition: tableau transformation}
in Theorem~\ref{theorem: tableau transformation} does not hold for $T'$.
Given a formula $\nu_1(p\Or q,a)$, the branch~$\branch{B}_0$ containing
only $\nu_1(p\Or q,a)$ is fully expanded.
The interpretation $\I(\branch{B}_0)$ constructed from~$\branch{B}_0$ as
defined in the previous section reflects the expressions $p$ and $q$.
The instantiation of the premise of the \eqref{rule: or}-rule 
with the expressions $p$ and $q$ belongs to the branch $\branch{B}_0$,
that is,
$\nu_1(p\Or q,a)\in\branch{B}_0$,
but $\I(\branch{B}_0)\not\models\nu_1(p,a)$ and $\I(\branch{B}_0)\not\models\nu_1(q,a)$.
This means condition~\eqref{condition: tableau transformation}
fails for~$\branch{B}_0$ and~$Y\define\{p,q\}$.

The following example shows that $T'$ is in fact incomplete.
Let $\branch{B}_1$ be the branch with formulae
$\nu_1(\Not p\Or\Not q,a), \nu_1(p,a), \nu_1(q,a)$.
The branch is fully expanded, because the
\eqref{rule: KE or}-rule is not applicable.
However the formulae are unsatisfiable.
This is why KE tableau calculi typically contain an analytic cut rule for completeness.

\section{Refinement based on Exploiting the Expressivity of the Logic}
\label{section: refinement 2}

In some cases, the object logic $L$ is expressive enough
to represent its own semantics.
For example, in the case of standard modal logics, any Kripke frame
condition can be encoded if a slightly more expressive hybrid modal
language is used~\cite{BlackburnDeRijkeVenema01,Blackburn-WHL-1998}.
This phenomenon leads us to consider a second kind of refinement,
where all `holds' predicates $\nu_1,\ldots,\nu_N$ and additional predicates 
of $\FO(\Lang{L})$ are expressible via validity of special 
expressions of the primary sort (concepts) of the object logic.

What does it mean for logic $L$ to be expressive enough to represent its
own semantics?
Assume that 
for every $n=0,\ldots,N$ and every $n$-ary predicate constant~$P$ occurring in the specification~$S_L$,
there are expressions 
\[
C_n^+(p,\ell_1,\ldots,\ell_n), \quad C_n^-(p,\ell_1,\ldots,\ell_n),
\quad
D_P^+(\ell_1,\ldots,\ell_n) \quad 
\text{and} \quad D_P^-(\ell_1,\ldots,\ell_n)
\]
of the primary sort (concepts),
depending on variable $p$ of sort $n$
and individual variables $\ell_1,\ldots,\ell_n$
of sort $0$,
such that the following all hold.
\begin{align}
\label{second_refinement_stmt_one}
&\forall S_L\models\forall x\left(\nu_1(C_n^+(p,\ell_1,\ldots,\ell_n),x)\rightarrow\nu_n(p,\nu_0(\ell_1),\ldots,\nu_0(\ell_n))\right)\\
&\forall S_L\models\forall x\left(\nu_1(C_n^-(p,\ell_1,\ldots,\ell_n),x)\rightarrow\Not\nu_n(p,\nu_0(\ell_1),\ldots,\nu_0(\ell_n))\right)\\
&\forall S_L\models\forall x\left(\nu_1(D_P^+(\ell_1,\ldots,\ell_n),x)\rightarrow P(\nu_0(\ell_1),\ldots,\nu_0(\ell_n))\right)\\
&\forall S_L\models\forall x\left(\nu_1(D_P^-(\ell_1,\ldots,\ell_n),x)\rightarrow \Not P(\nu_0(\ell_1),\ldots,\nu_0(\ell_n))\right)
\label{second_refinement_stmt_four}
\end{align}
It is worth noting that because the equality theory is included in the
specification~$S_L$
the following also hold:
\begin{align*}
&\forall S_L\models\forall x\left(\nu_1(D_\approx^+(\ell_1,\ell_2),x)\rightarrow \nu_0(\ell_1)\approx\nu_0(\ell_2)\right),\\
&\forall S_L\models\forall x\left(\nu_1(D_\approx^-(\ell_1,\ell_2),x)\rightarrow \nu_0(\ell_1)\not\approx\nu_0(\ell_2)\right).
\end{align*}\medskip

\noindent If there are expressions such that
(\ref{second_refinement_stmt_one})--(\ref{second_refinement_stmt_four}) 
are true it is possible to express all tableau rules in~$T_L$
in the object language $\Lang{L}$ itself as follows.

Let $\varepsilon$ be a one-to-one mapping of domain variables
to variables of sort~$0$.
Now we only need to  
replace 
every positive occurrence of $\nu_n(E,x_1,\ldots,x_n)$ in~$T_L$
with the concept $C_n^+(E,\varepsilon(x_1),\ldots,\varepsilon(x_n))$,
every (negative) occurrence of $\Not\nu_n(E,x_1,\ldots,x_n)$ in~$T_L$
with the concept $C_n^-(E,\varepsilon(x_1),\ldots,\varepsilon(x_n))$.
Similarly, all predicate constants~$P$ need to be replaced
with occurrences of $D_P^+$ or $D_P^-$ depending on the polarity of~$P$.
Then the domain sort $N+1$ of the meta-language
$\FO(\Lang{L})$ is reflected by the sort $0$.

A small technical complication is caused by  
functions in $\FO(\Lang{L})$ (Skolem functions and Skolem constants, in particular)
occurring in the generated tableau rules.
For them there may not be corresponding function symbols in the object language~$\Lang{L}$.
This can be addressed by introducing new connectives
$f_g$ into~$\Lang{L}$ 
for every  
function~$g$ (including 
constants)
of $\FO(\Lang{L})$ so
that for any $p_1,\ldots,p_m,\ell_1,\ldots,\ell_n$,
the term
 $f_g(p_1,\ldots,p_m,\ell_1,\ldots,\ell_n)$ is of sort $0$
and its semantics is defined by
 \[\nu_0(f_g(p_1,\ldots,p_m,\ell_1,\ldots,\ell_n))\define
 g(p_1,\ldots,p_m,\nu_0(\ell_1),\ldots,\nu_0(\ell_n)).\]
An alternative is to introduce unique, new
individual constants
(for every $p_1,\ldots,p_m$, $\ell_1,\ldots,\ell_n$)
instead of new connectives. 

If $T$ is a tableau calculus for the logic $L$
we denote by $\Rfe(T)$ the \emph{refined tableau calculus}
obtained by replacing every positive occurrence of $\nu_n(E,x_1,\ldots,x_n)$ in~$T_L$
by the concept $C_n^+(E,\varepsilon(x_1),\ldots,\varepsilon(x_n))$,
every occurrence of $\Not\nu_n(E,x_1,\ldots,x_n)$
by $C_n^-(E,\varepsilon(x_1),\ldots,\varepsilon(x_n))$,
every positive occurrence of a predicate constants~$P$ by
$D_P^+$,
every negative occurrence of a predicate constants~$P$ by
$D_P^-$,
and every  
function $g$ with the new connective
$f_g$.

\begin{theorem}
\label{theorem: rfe}
Let $T$ be a sound and complete tableau calculus for a logic $L$.
If there are expressions such that
(\ref{second_refinement_stmt_one})--(\ref{second_refinement_stmt_four}) 
then $\Rfe(T)$ is sound and complete.
If, in addition, $T$ is constructively complete
then $\Rfe(T)$ is also constructively complete for $L$.
\end{theorem}

To illustrate the refinement introduced in this section we enrich the object language
of \SO with an additional connective.
In particular, we add the colon connective $:$, with sort~$(0,1,1)$,
defined by:
\[
\forall x\;\bigl(\nu_1(\ell:p,x)\equiv%
 \nu_1(p,\nu_0(\ell))\bigr).
\]
We also introduce connectives which correspond to Skolem functions
into the object language.

This allows us to find object expressions for defining the predicates
$\approx$, $\nu_1$ and~$\nu_2$ in the language of the
logic: 
\begin{align*}
  C_1^+(p,\ell)&\define\ell:p, & C_1^-(p,\ell)&\define\ell:\Not p,\\
  C_2^+(r,\ell_1,\ell_2)&\define\ell_1:\exists r.\{\ell_2\}, & C_2^-(r,\ell_1,\ell_2)&\define\ell_1:\Not\exists r.\{\ell_2\},\\
  D_\approx^+(\ell_1,\ell_2)&\define\ell_1:\{\ell_2\}, & D_\approx^-(\ell_1,\ell_2)&\define\ell_1:\Not\{\ell_2\}.
 \end{align*}\medskip

\noindent This means the notation of the tableau calculus can be refined and simplified.
The refined and simplified rules are given in Figure~\ref{fig: optimised SO tableau}.
Comparing Figure~\ref{fig: SO tableau} and Figure~\ref{fig: optimised SO tableau}
we can see that the refined formulations of the rules 
\begin{gather*}
\begin{alignedat}{1}
&\tableaurule{\nu_1(\{\ell\},x)}{\nu_0(\ell)\approx x}%
&\qquad\quad
&\tableaurule{\Not\nu_1(\{\ell\},x)}{\nu_0(\ell)\not\approx x}%
&\qquad\quad
&\tableaurule{\nu_1(\Not p,x)}{\Not\nu_1(p,x)}%
\end{alignedat}
\end{gather*}
are all redundant and can be removed from the refined tableau calculus
since their premises coincide with the conclusions.
Furthermore,
the refined equality congruence rules equivalently reduce to a smaller set of rules.
For instance, the refined rule of transitivity of the equality
\[
\tableaurule{\ell:\{\ell'\}\tand \ell':\{\ell''\}}{\ell:\{\ell''\}} 
\]
can be derived from the following rules.
\begin{gather*}
\begin{alignedat}{1}
&  \tableaurule{\vphantom{Pp}\ell:\{\ell'\}}{\vphantom{Pp}\ell':\{\ell\}}
&\qquad\quad
&  \tableaurule{\vphantom{Pp}\ell:p\tand \ell:\{\ell'\}}{\vphantom{Pp}\ell':p}
\end{alignedat}
\end{gather*}
Finally, the closure rule for equality is subsumed by the usual closure rule.

By Theorems~\ref{theorem: tableau transformation} 
and~\ref{theorem: rfe} the rules
in Figure~\ref{fig: optimised SO tableau} provide a sound and
(constructively) complete labelled tableau calculus for the
logic~$\SO$.

\begin{figure}[!tb]
\begin{trivlist}
\item
Decomposition rules:
\begin{gather*}
\begin{alignedat}{1}
&\tableaurule{\ell:\Not\Not p}{\ell:p}%
&\qquad
&\tableaurule{\ell:(p\Or q)}{\ell:p\tor\ell:q}
&\qquad%
&\tableaurule{\ell:\Not(p\Or q)}{\ell:\Not p\tand\ell:\Not q}
\end{alignedat}
\\
\begin{alignedat}{1}
\tableaurule{\ell:\exists r.p}{\ell:\exists r.\{f(r,p,\ell)\}\tand f(r,p,\ell):p}
&\qquad
\tableaurule{\ell:\Not\exists r.p\tand\ell:\exists r.\{\ell'\}}%
                {\ell':\Not p}
\end{alignedat}
\end{gather*}
\item
Transitivity rule:
\begin{gather*}
\tableaurule{\ell:\exists r.\{\ell'\}\tand \ell':\exists r.\{\ell''\}}{\ell:\exists r.\{\ell''\}}
\end{gather*}
\item
Equality congruence rules:
\begin{gather*}
  \tableaurule{\vphantom{Pp}\ell:\{\ell'\}}{\vphantom{Pp}\ell':\{\ell\}}
  \qquad
  \tableaurule{\vphantom{Pp}\ell:\Not\{\ell'\}}{\vphantom{Pp}\ell':\{\ell'\}}
  \qquad
  \tableaurule{\vphantom{Pp}\ell:p}{\vphantom{Pp}\ell:\{\ell\}}
  \qquad
  \tableaurule{\vphantom{Pp}\ell:\Not\exists r.\{\ell'\}}{\vphantom{Pp}\ell':\{\ell'\}}
  \qquad 
  \tableaurule{\vphantom{Pp}\ell:p\tand \ell:\{\ell'\}}{\vphantom{Pp}\ell':p}
\\
  \tableaurule{\vphantom{Pp}\ell:\exists r.\{\ell'\}\tand \ell':\{\ell''\}}{\vphantom{Pp}\ell:\exists r.\{\ell''\}}
  \qquad
  \tableaurule{\vphantom{Pp} f(r,p,\ell):\{f(r,p,\ell)\}\tand \ell:\{\ell'\}}%
           {\vphantom{Pp}f(r,p,\ell):\{f(r,p,\ell')\}}
\end{gather*}
\item
Closure rule:
\begin{gather*}
\tableaurule{\ell:p\tand\ell:\Not p}{\bot}%
\end{gather*}
\end{trivlist}
\caption{Refined tableau rules for \SO.}\label{fig: optimised SO tableau}
\end{figure}

\section{Termination through Unrestricted Blocking}\label{section: blocking}

We say a tableau calculus $T$ is
\emph{terminating} (for satisfiability) iff for every \emph{finite}
set of concepts~$\mathcal{S}$ every closed tableau~$T(\mathcal{S})$
is finite and every open tableau~$T(\mathcal{S})$ has a finite
open branch.

For some logics, for example, modal logic $\ml{K}$,
the synthesised
tableau calculi are terminating but in general they are not.
In order to guarantee termination, various blocking 
mechanisms have been developed.
Generally one can distinguish between at least three kinds of blocking techniques:
those that reuse domain terms, those that are based on case
analysis over conjectured equality constraints between domain terms and
equality reasoning, and specialised loop checking mechanisms.
Approaches based on reusing domain terms have been used
for minimal model generation for classical logic~\cite{BryManthey87,BryTorge98}. 
Approaches based on conjectured equality constraints include
\cite{BaumgartnerSchmidt06,HustadtSchmidt99b,SchmidtTishkovsky-UTD+-2007}.
Loop checking mechanisms are based on comparing sets of concepts (expressions
of sort~$1$) labelled by the same domain terms (or individuals)
with minimal equality reasoning and without explicitly conjectured
equality constraints and backtracking.
Several such loop checking mechanisms have been developed for different
modal and description logics, but also hybrid logics and other
logics~\cite{HughesCresswell68,BaaderSattler01,HorrocksSattler-TDP_SHOIQ-2007,BolanderBlackburn07,CialdeaMayerCerrito01}.

In this section we adopt the unrestricted blocking mechanism
of~\cite{SchmidtTishkovsky-UTD+-2007} to obtain terminating tableau
calculi.
An alternative that could also be used is blocking through reusing
domain terms, but this would have required changing the rules of the
calculus.
Both unrestricted blocking and blocking through reuse of terms are
less specialised and more generic than standard loop checking
mechanisms.

Though introduced for deciding expressive description logics with role
negation, the applicability of the unrestricted blocking mechanism is
not limited to description logics~\cite{SchmidtTishkovsky-GTM+-2008}.
It provides a powerful method for obtaining tableau decision procedures
for logics with the effective finite model property (with respect to
their semantics).

A logic~$L$ has the \emph{effective finite model property}
iff there is a computable function $\mu$,
 with the set of all finite sets of concept expressions as domain and
 a subset of the set of natural numbers as range, 
 such that the following holds:
           For every finite set of concept expressions $\mathcal{S}$,  
           if $\mathcal{S}$ is satisfiable in an $L$-model
           then there is a finite $L$-model for $\mathcal{S}$ 
           with the number of elements in the domain not exceeding $\mu(\mathcal{S})$.

The unrestricted blocking mechanism is based on adding the following
rule, called the \emph{unrestricted blocking} rule, to a sound and
complete tableau calculus.
\[
\tableaurule{x\approx x\tand y\approx y}{x\approx y\tor x\not\approx y}[ub]
\label{rule: unrestricted blocking}
\]
In our context the idea is that the rule conjectures whether pairs
of domain terms (of sort~$N+1$ in $\FO(\Lang{L})$) on the current
branch are equal or not.
In the left branch two such terms are conjectured to be equal.
If this leads to a contradiction then they cannot be equal, which is
the information carried by the right branch.
The rule is generally sound, thus adding it to any sound and
(constructive) complete tableau calculus preserves soundness and
(constructive) completeness.

For termination it is crucial to impose additional restrictions
on the application of the rules in the tableau calculus that
introduce new domain terms to the derivation.
This is achieved by defining an ordering $<$ on terms and
imposing conditions~(c\ref{req: blocking}) and~(c\ref{req: ub before exists})
below on the calculus.

In particular, let $<$ be an ordering of terms of the domain sort $N+1$
in the branch which is a linear extension of the order of the introduction
of the terms during the derivation.
That is, $t<t'$, whenever the first appearance of term $t'$ in the
branch is strictly later than the first appearance of term $t$.
The mentioned conditions are:
\begin{enumerate}[(c1)]
\item\label{req: blocking}
If $t\approx t'$ appears in a branch and $t<t'$, then
possible applications of any rules to formulae with the term $t'$
producing new terms of the domain sort
are not performed.
\item\label{req: ub before exists}
In every open branch there is some node from which point onwards
before any application of any rules which produce new terms of the domain sort
all possible applications of the \eqref{rule: unrestricted blocking}~rule
have been performed. 
\end{enumerate}\medskip

\noindent Condition~(c\ref{req: blocking}) specifies that term-producing rules
may only be applied to formulae where the domain terms are the smallest
representatives in their equivalence classes.
The positive rule for $\exists\cdot.\cdot$ is
the only term-producing rule in the calculus for $\SO$.
Condition~(c\ref{req: ub before exists}) says that at some point in a branch
the unrestricted blocking rule has been applied exhaustively before
the application of term-producing rules.

For a tableau calculus $T$ we denote by $T+\eqref{rule: unrestricted blocking}$
a tableau calculus obtained from $T$ by adding the above blocking mechanism based
on the unrestricted blocking rule.

According to~\cite{SchmidtTishkovsky-GTM+-2008},
one of the prerequisites for termination of the calculus $T+\eqref{rule: unrestricted blocking}$
is the subexpression property of $T$.
Let $\preceq$ be a reflexive and transitive ordering on $\Lang{L}$-expressions.
Following~\cite{SchmidtTishkovsky-GTM+-2008},
we say that a tableau calculus $T$ is \emph{compatible with~$\sub_\preceq$},
or has the \emph{subexpression property} with respect to $\preceq$,
iff for every set of concepts~$\mathcal{S}$,
all $\Lang{L}$-expressions occurring in the tableau derivation~$T(\mathcal{S})$
belong to $\sub_\preceq(\mathcal{S})$.

Given a well-defined semantic specification $S$ 
the process of construction of $T$ from~$S$ described in
Section~\ref{section_tableau_synthesis}
ensures that 
every rule
of $T$ is monotone with respect to the ordering~$\prec$ 
induced by $S$.
That is, every $\Lang{L}$-expression in
each conclusion of a rule is not greater with respect to $\prec$
than $\Lang{L}$-expressions in the premises of the rule.
Therefore, we can conclude that~$T$
has the subexpression property with respect to the reflexive closure of the ordering~$\prec$.
Thus: 
\begin{lemma}\label{lemma: subexpresiion property}
Let $\preceq$ be a reflexive closure of the ordering $\prec$
induced by a well-defined semantic specification $S$.
Then the tableau calculus~$T$ 
generated from $S$
has the subexpression property with respect to $\preceq$.
\end{lemma}

This property is a necessary condition for termination of the calculus enhanced
by the unrestricted blocking rule
mechanism~\cite{SchmidtTishkovsky-UTD+-2007,SchmidtTishkovsky-GTM+-2008}.
Another necessary condition for termination is
finiteness of $\sub_\preceq$. 
The operator $\sub$ mapping sets of concepts to sets of expressions 
is \emph{finite} iff
$\sub(\mathcal{S})$ is finite for every finite set of concepts $\mathcal{S}$.
By K\"onig's Infinity Lemma, $\sub_\preceq$ is finite 
whenever~$\prec$~is well-founded and finitely branching.
Therefore:
\begin{lemma}\label{lemma: sub is finite}
Let $\preceq$ be a reflexive closure of the ordering $\prec$
induced by a well-defined semantic specification $S$.
If $S^+\cup S^-$ is finite
then the operator $\sub_\preceq$ is finite.
\end{lemma}

Reformulating the main result in~\cite{SchmidtTishkovsky-GTM+-2008}
in terms of the notation of this paper gives us:
\begin{theorem}\label{theorem: termination}
\label{theorem_general_termination}
Let $L$ be a logic and $T$ be a sound and constructively complete tableau calculus for a semantic
specification $S_L$ of the logic $L$.
Then $T+\eqref{rule: unrestricted blocking}$ is sound and constructively complete for $S_L$.
Furthermore, $T+\eqref{rule: unrestricted blocking}$ is
terminating for $L$, if the following conditions all hold:
\begin{enumerate}[\em(1)]
\item
There is a finite closure operator $\sub$ (defined on sets of concepts
of the language of~$L$) such that
$T$ is compatible with 
$\sub$.
\item
$L$ has the effective finite model property with respect to $S_L$.
\end{enumerate}
\end{theorem}\medskip

\noindent From this theorem and Theorems~\ref{theorem_soundness}, \ref{theorem_constructive_completeness},
\ref{theorem: tableau transformation}, \ref{theorem: rfe} and Lemmas~\ref{lemma: subexpresiion property}
and~\ref{lemma: sub is finite} it follows
that the extensions of the generated and refined tableau calculi 
with unrestricted blocking
are sound and (constructively) complete.
Moreover, if it is known that the given logic has the effective finite
model property with respect to a finite semantic specification then both
extensions are terminating as well.

It is well known that $\SO$
has the effective finite model property
with respect to $S_\SO$, and clearly $S_\SO$ has a finite number of statements.
As a consequence, a terminating tableau calculus for \SO is obtained if the
calculus in Figure~\ref{fig: optimised SO tableau} 
is enhanced with the unrestricted blocking mechanism as described above.
Using the refinements in Section~\ref{section: refinement 2}
the unrestricted blocking rule can be transformed as follows.
\[
\tableaurule{\ell:\{\ell\}\tand\ell':\{\ell'\}}{\ell:\{\ell'\}\tor\ell:\Not\{\ell'\}}[ub$'$] \label{rule: SO unrestricted blocking}
\]\medskip

\noindent Let $\mathcal{T}_\SO$ be a tableau calculus comprising of the rules listed in Figure~\ref{fig: optimised SO tableau}
and the rule~\eqref{rule: SO unrestricted blocking}.

\begin{theorem}
The calculus $\mathcal{T}_\SO$ is sound and constructively complete for $\SO$.
Furthermore, $\mathcal{T}_\SO$ is terminating provided that conditions~(c\ref{req: blocking})
and~(c\ref{req: ub before exists}) are both true for $\mathcal{T}_\SO$-derivations.
\end{theorem}
In the calculus $\mathcal{T}_\SO$, the imposed conditions~(c\ref{req: blocking}) 
and~(c\ref{req: ub before exists}) are
restrictions on applications of the rule
\[\tableaurule{\ell:\exists r.p}{\ell:\exists r.\{f(r,p,\ell)\}\tand f(r,p,\ell):p}.\]
Following~\cite{SchmidtTishkovsky-GTM+-2008},
the calculus $\mathcal{T}_\SO$ can be turned into a deterministic decision procedure
using breadth-first search or depth-first search.

The calculus $\mathcal{T}_\SO$
presents a new terminating tableau calculus for \SO or equivalent
hybrid logics.
The main difference to existing tableau approaches (in a similar
style) for~\SO or equivalent hybrid logics is that the individuals
(or nominals) are handled differently. To force termination typically
either equality or subset ancestor loop checking is used, and often
transitivity is handled by a propagation rule.

\section{Synthesising Tableau Calculi for Intuitionistic Logic}
\label{section_case_studies}

We consider another example to illustrate the method.
Propositional intuitionistic logic~$\IPC$ is a logic
where the `holds' predicates $\nu_1,\ldots,\nu_N$ cannot be expressed in the language of the logic.
It is non-Boolean and provides an example of a logic 
where the background theory
interacts with the definitions of the connectives.

The language of intuitionistic logic is a one-sorted language defined
over a countable set of propositional symbols $p,q,p_1,p_2,\ldots$,
and the standard connectives are $\Imp, \lOr, \lAnd, \bot$.
The semantic specification $S_\IPC$ in $\FO(\Lang{L})$ is given by (confer~\cite{Kripke-SAIL1-1965}):
\begin{trivlist}
\item[\hskip\parindent] Connective definitions:
\[\setlength{\arraycolsep}{1pt}
\begin{array}{lrl}
&\forall x\;\bigl(\nu_1(\bot,x) &\equiv\bot\bigr)\\
&\forall x\;\bigl(\nu_1(p\lAnd q,x)&\equiv%
\nu_1(p,x)\lAnd\nu_1(q,x)\bigr)\\
&\forall x\;\bigl(\nu_1(p\lOr q,x)&\equiv%
\nu_1(p,x)\lOr\nu_1(q,x)\bigr)\\
&\forall x\;\bigl(\nu_1(p\Imp q,x)&\equiv%
\forall y\;\bigl(R(x,y)\Imp(\nu_1(p,y)\Imp\nu_1(q,y)\bigr)\bigr)
\end{array}
\]
\item[\hskip\parindent] Background theory:
\[\setlength{\arraycolsep}{1pt}
\begin{array}{lrl}
&\forall x& R(x,x)\\
&\forall x\forall y&(R(x,y)\lAnd R(y,x)\Imp x\approx y)\\
&\forall x\forall y\forall z&(R(x,y)\lAnd R(y,z)\Imp R(x,z))\\
&\forall x\forall y&\bigl(\nu_1(p,x)\lAnd R(x,y)\Imp\nu_1(p,y)\bigr)
\end{array}
\]
\end{trivlist}
The connective definitions impose
the usual requirements for truth of a formula
in a world of an intuitionistic Kripke model.
For instance, the definition of implication
expresses in~$\FO(\Lang{L})$ the property
that an implication of $q$ from $p$ is true in a world $x$
if and only if $q$ is true in every successor of $x$
whenever $p$ is true in that successor.
$R$~is the domain predicate symbol representing a partial order, which
is specified by the first three sentences of the background theory.
The last sentence in the background theory specifies
monotonicity of the truth of formulae (of sort~$1$).

For intuitionistic logic the orderings $\prec_0$ and $\prec$
coincide.
The ordering $\prec$ on subexpressions induced by the semantic
definition of the connectives
is the smallest ordering satisfying
$E_1\prec E_1\sigma E_2$ and $E_2\prec E_1\sigma E_2$, 
for each $\sigma\in\{\Imp,\lOr,\lAnd\}$
and any intuitionistic formulae $E_1$ and $E_2$.
That is, $\prec$ is the direct subexpression ordering on intuitionistic formulae.
Thus, the closure operator $\sub_\preceq$ induced by the reflexive closure $\preceq$
of the ordering $\prec$ is finite.

\begin{figure}[tb]
\begin{trivlist}
\item
Decomposition rules:
\begin{gather*}
\begin{alignedat}{1}
&\tableaurule{\nu_1(\bot,x)}{\bot}%
&\qquad\quad
&\tableaurule{\neg\nu_1(\bot,x)}{\neg \bot}%
\end{alignedat}
\\
\begin{alignedat}{1}
&\tableaurule{\nu_1(p\lAnd q,x)}{\nu_1(p,x)\tand\nu_1(q,x)}%
&\qquad\quad%
&\tableaurule{\Not\nu_1(p\lAnd q,x)}{\Not\nu_1(p,x)\tor\Not\nu_1(q,x)}%
\end{alignedat}
\\
\begin{alignedat}{1}
&\tableaurule{\nu_1(p\lOr q,x)}{\nu_1(p,x)\tor\nu_1(q,x)}%
&\qquad\quad%
&\tableaurule{\Not\nu_1(p\lOr q,x)}{\Not\nu_1(p,x)\tand\Not\nu_1(q,x)}%
\end{alignedat}
\\
\tableaurule{\nu_1(p\Imp q,x) \tand y\approx y}%
                {\neg R(x,y)\tor\neg\nu_1(p,y)\tor\nu_1(q,y)}%
\\
\tableaurule{\Not\nu_1(p\Imp q,x)}{R(x,f(p,q,x))\tand\nu_1(p,f(p,q,x))\tand\Not\nu_1(q,f(p,q,x))}%
\end{gather*}
\item
Theory rules:
\begin{gather*}
\tableaurule{x\approx x}{R(x,x)}%
\qquad\quad%
\tableaurule{x\approx x \tand y\approx y}{\Not R(x,y)\tor\Not R(y,x)\tor x\approx y}%
\qquad\quad%
\tableaurule{x\approx x \tand y\approx y \tand z\approx z}{\Not R(x,y)\tor\Not R(y,z)\tor R(x,z)}%
\\
\tableaurule{p\approx p\tand x\approx x \tand y\approx y}{\Not\nu_1(p,x)\tor \Not R(x,y)\tor \nu_1(p,y)}%
\end{gather*}
\item
Closure rules:
\begin{gather*}
\begin{alignedat}{1}
&\tableaurule{\nu_1(p,x)\tand\Not\nu_1(p,x)}{\bot}%
&\qquad\quad%
&\tableaurule{R(x,y)\tand\Not R(x,y)}{\bot}%
\end{alignedat}
\end{gather*}
\end{trivlist}
\caption{Generated tableau rules for intuitionistic logic.}\label{fig: int tableau}
\end{figure}

The tableau rules generated from the specification $S_\IPC$ are those listed in
Figure~\ref{fig: int tableau}.
Together with the equality rules of Figure~\ref{fig: equality rules}, they form a calculus, which is
sound and constructively complete 
for intuitionistic logic.
This is a consequence of
Theorems~\ref{theorem_soundness}
and~\ref{theorem_constructive_completeness}.

Refining the generated rules
yields the rules listed in
Figure~\ref{fig: optimised int tableau}.
Using Theorem~\ref{theorem: tableau transformation}
we conclude that together with the equality rules these rules provide a sound and
constructively complete tableau calculus for intuitionistic logic.
We denote this calculus by $T_\IPC$.

\begin{figure}[tb]
\begin{trivlist}
\item
Decomposition rules:
\begin{gather*}
\begin{alignedat}{1}
&\tableaurule{\nu_1(\bot,x)}{\bot}%
&\qquad\quad
&\tableaurule{\nu_1(p\lAnd q,x)}{\nu_1(p,x)\tand\nu_1(q,x)}%
&\qquad\quad%
&\tableaurule{\Not\nu_1(p\lAnd q,x)}{\Not\nu_1(p,x)\tor\Not\nu_1(q,x)}%
\end{alignedat}
\\
\begin{alignedat}{1}
&&\tableaurule{\nu_1(p\lOr q,x)}{\nu_1(p,x)\tor\nu_1(q,x)}%
&\qquad\quad%
&\tableaurule{\Not\nu_1(p\lOr q,x)}{\Not\nu_1(p,x)\tand\Not\nu_1(q,x)}%
&\qquad\quad
&\tableaurule{\nu_1(p\Imp q,x)\tand R(x,y)}%
                {\Not\nu_1(p,y)\tor\nu_1(q,y)}%
\end{alignedat}
\\
\tableaurule{\Not\nu_1(p\Imp q,x)}{R(x,f(p,q,x))\tand\nu_1(p,f(p,q,x))\tand\Not\nu_1(q,f(p,q,x))}%
\end{gather*}
\item
Theory rules:
\begin{gather*}
\begin{alignedat}{3}
&\tableaurule{x\approx x}{R(x,x)}%
&\qquad\quad%
&\tableaurule{R(x,y)\tand R(y,x)}{x\approx y}%
&\qquad\quad%
&\tableaurule{R(x,y)\tand R(y,z)}{R(x,z)}%
&\qquad\quad%
&\tableaurule{\nu_1(p,x)\tand R(x,y)}{\nu_1(p,y)}%
\end{alignedat}
\end{gather*}
\item
Closure rules:
\begin{gather*}
\begin{alignedat}{1}
&\tableaurule{\nu_1(p,x)\tand\Not\nu_1(p,x)}{\bot}%
\end{alignedat}
\end{gather*}
\end{trivlist}
\caption{Refined tableau rules for intuitionistic logic.}\label{fig: optimised int tableau}
\end{figure}

Similarly to the case of \SO, because intuitionistic logic
has the effective finite model property,
by Theorem~\ref{theorem: termination} together with Lemmas~\ref{lemma: subexpresiion property} and~\ref{lemma: sub is finite},
a terminating tableau calculus for $\IPC$ 
is obtained if the calculus $T_\IPC$ is enhanced with the unrestricted blocking mechanism.

\begin{theorem}
The tableau calculus $T_\IPC+\eqref{rule: unrestricted blocking}$
is sound, constructively complete and terminating
for $\IPC$.
\end{theorem}

Following~\cite{SchmidtTishkovsky-GTM+-2008}, $T_\IPC+\eqref{rule: unrestricted blocking}$ can be turned into deterministic decision
procedures for $\IPC$ using breadth-first search or depth-first
search.

\section{Discussion and Conclusions}
\label{section_conclusion}

The method introduced in this paper automatically produces a sound and
constructively complete tableau calculus from the semantic first-order
specification of a many-sorted logic.
The method is directly applicable to many non-classical logics
and covers many types of ground tableau calculi commonly found in the
literature.

On one hand, the formalisation is based on ideas used
in the implementation
of tableau decision procedures for
modal and description logics in 
the \mettel system~\cite{mettel_page,TishkovskySchmidtKhodadadi11}.
The \mettel system provides a core for tableau derivations, which does
not depend on a logical language.
Due to this language flexibility, without any  
modification of the core code, the prover constructs (sound, complete,
and terminating) tableau derivations for standard modal
logics, superintuitionistic logics (via the G\"odel translation), many
description logics,  
as well as for logics of metrics and topology
for which it was originally written.
Termination is achieved via an implementation 
of generalisations of standard blocking mechanisms
as well as the unrestricted blocking mechanism.
This means that \mettel provides an implementation
of a tableau
decision procedure for description logics with full support of the
role negation operator, which can not currently be handled by other
tableau-based description and modal logic theorem provers.
On the other hand, the results of this paper provide
the theoretical foundation for the correct behaviour of tableau algorithms
implemented in \mettel. 

More importantly, the results
can be viewed
as providing a mathematical formalisation
and generalisation
of tableau development methodologies.
The formalisation separates
the creative part of tableau calculus development,
which needs to be done by a human developer,
and the automatic part of the development process,
which can be left to an automated (currently first-order) prover
and an automated tableau synthesiser.
In general, there is no algorithm for checking that an arbitrarily given binary relation
forms a well-founded ordering.
Therefore the creative part is writing down the semantic specification
of the object logic so that the conditions of well-foundedness of the
orderings~$\prec_0$ and $\prec$ hold.
The automatic part deals with verification
of the first-order conditions~\ref{condition: well-defined semantics 1}
and~\ref{condition: well-defined semantics 3a},
and the generation of tableau rules from the (well-defined) semantics
provided by the developer.

For common modal and description logics
conditions~\ref{condition: well-defined semantics 1}
and~\ref{condition: well-defined semantics 3} are simple to
check, 
even trivial in many cases.
In fact, a developer usually implicitly formalises the 
logic's semantics $S$ in such a way that $S=S^0\cup
S^b$. 
This is the case for almost all known logics.
If the specification of the semantics satisfies $S=S^0\cup
S^b$
then conditions~\ref{condition: well-defined semantics 1} and~\ref{condition: well-defined semantics 3} hold trivially and the orderings 
$\prec_0$ and $\prec$ coincide.
This means the ordering used for the specification of the semantics
of the logical connectives (which is usually well-founded) is enough
for tableau synthesis.

The following are examples of first-order
definable logics, which all have a normalised and well-defined semantic specification according to
the definitions in Section~\ref{section_semantics}: 
 \begin{enumerate}[$\bullet$]
  \item most description logics, including \ALCO, \SO, \ALBO~\cite{SchmidtTishkovsky-UTD+-2007}, \SHOIQ~\cite{HorrocksSattler-TDP_SHOIQ-2007};
  \item most propositional modal logics, including \ml{K}, \KFour, \ml{S4}, \ml{KD45}, \SFive;
  \item propositional intuitionistic logic~\cite{Kripke-SAIL1-1965} and 
     many Kripke-complete propositional superintuitionistic logics;
  \item the logic of metric and topology~\cite{HustadtTishkovskyWolterZakharyaschev-ARMT-2006}.
 \end{enumerate}\medskip

\noindent This paper
also presents a general
method for proving (constructive) completeness of tableau calculi.
In addition, the generated rules can be transformed to 
the rules with lower branching factors
provided that
condition~\eqref{condition: tableau transformation} has been proved
by induction on the ordering~$\prec$ for the refined calculus.

With enough expressivity for representing the basics of the semantics 
within the logic it is possible to simplify the language of the tableau calculus.
In this case, the obtained calculus is similar to tableau
calculi for description logics with singleton concepts,
but also hybrid modal
logic~\cite{BolanderBlackburn07} and labelled
tableau calculi~\cite{Fitting-PMMIL-1983,Gore-TMMTL-1999}.
Otherwise, the calculus has the same flavour as standard
tableau calculi for intuitionistic logic, where
every node in a tableau derivation is characterised by two complementary sets of true and
false formulae (concepts).

That the generated calculi are constructively complete has the added
advantage that models can be effectively generated from open, finished
branches in tableau derivations.
This means that the synthesised tableau calculi can be used for
finding models.
If the calculus includes the unrestricted blocking mechanism
various strategies on the application of the unrestricted blocking rule
can be employed for obtaining models with minimal domain
sizes.

As case studies
we considered tableau synthesis for
propositional intuitionistic logic and the description logic \SO
with singleton concepts and transitive roles. 
We believe the approach is also applicable to most known, first-order
definable modal and description 
logics including the ones mentioned above. Non first-order translatable logics such as propositional dynamic logic
are currently beyond the scope of the method.

The tableau calculi generated are Smullyan-type ta\-b\-leau calculi,
that is, ground semantic tableau calculi.
We believe that other types of tableau calculi can be
generated using the same techniques.
We expect that generating unlabelled tableau calculi without
explicit background predicates or domain terms will be
possible, at least to some extent, but this is not immediate.
One possibility would be to investigate if these can be derived as
further refinements of the labelled tableau calculi generated by
method presented in this paper.
Such a line of investigation would be interesting and shed
more light on the relationship between different kinds of tableau
calculi. 
Exploiting the known relationships to other deduction methods we
expect synthesis of non-tableau approaches is possible as well,
but all this is future work.

Further investigations are needed to explore the extension of the
framework to generate calculi based on propagation rules which
incorporate frame correspondence properties into the definition
of connectives to replace the theory rules for modal and description
logics (for example, transitivity for the logic \SO).
It is clear though that this is a much harder problem because
guaranteeing completeness becomes more difficult.
It is also clear that no results at the same level of generality as for
the use of theory rules in this paper can be expected.

A future goal is to further reduce human involvement in the development
of calculi by finding appropriate automatically
verifiable conditions for refined calculi to be
generated.

We plan to implement the methodology as an automatic generator
of tableau calculi.
This will give users the ability to obtain tableau calculi very easily
and without needing to have relevant knowledge of tableau-based reasoning
or experience in developing tableau calculi.
Combined with a prover engineering platform
such as \lotrec~\cite{GasquetEtAl05} or the Tableau
Workbench~\cite{AbateGore03} there is even the potential to build
systems that would allow users
to get implemented provers from the specification of logics.
\lotrec and the Tableau Workbench are generic systems for building
tableau-based theorem provers for non-classical logics.
Currently they allow users to define tableau procedures by
flexibly specifying the set of tableau rules, the search strategies,
the blocking technique and the optimisation techniques to be used.
This is then compiled into a specialised prover for the specified
procedure.
Enhanced with the tableau synthesis methodology, such systems could 
allow the user to define just the logic and produce an
implemented prover for this logic.

\end{document}